\documentclass[twocolumn]{aastex63}
\usepackage{comment}

\newcommand{\muJyb}{$\ \rm \mu Jy\ beam^{-1}$}

\shorttitle{A Circumplanetary Disk Around PDS70\,c}
\shortauthors{Benisty et al.}

\graphicspath{{./}{FIG/}}

\begin{document}
\title{A Circumplanetary Disk Around PDS70\,c}

\correspondingauthor{Myriam Benisty}
\email{Myriam.Benisty@univ-grenoble-alpes.fr}

\author[0000-0002-7695-7605]{Myriam Benisty}
\affiliation{Unidad Mixta Internacional Franco-Chilena de Astronom\'ia, CNRS, UMI 3386. Departamento de Astronom\'ia, Universidad de Chile, Camino El Observatorio 1515, Las Condes, Santiago, Chile} 
\affiliation{Univ. Grenoble Alpes, CNRS, IPAG, 38000 Grenoble, France}
\author[0000-0001-7258-770X]{Jaehan Bae}
\altaffiliation{NASA Hubble Fellowship Program Sagan Fellow}
\affil{Earth and Planets Laboratory, Carnegie Institution for Science, 5241 Broad Branch Road NW, Washington, DC 20015, USA}
\author[0000-0003-4689-2684]{Stefano Facchini}
\affiliation{European Southern Observatory, Karl-Schwarzschild-Str. 2, 85748 Garching, Germany}
\author[0000-0001-7250-074X]{Miriam Keppler}
\affiliation{Max Planck Institute for Astronomy, K\"onigstuhl 17, 69117, Heidelberg, Germany}
\author[0000-0003-1534-5186]{Richard Teague}
\affiliation{Center for Astrophysics \textbar\, Harvard \& Smithsonian, 60 Garden St., Cambridge, MA 02138, USA}
\author[0000-0002-0786-7307]{Andrea Isella}
\affiliation{Department of Physics and Astronomy, Rice University, 6100 Main Street, MS-108, Houston, TX 77005, USA}
\author[0000-0002-2358-4796]{Nicolas T. Kurtovic}
\affiliation{Max Planck Institute for Astronomy, K\"onigstuhl 17, 69117, Heidelberg, Germany}
\author[0000-0002-1199-9564]{Laura M. Pérez}
\affiliation{Departamento de Astronom\'a, Universidad de Chile, Camino El Observatorio 1515, Las Condes, Santiago, Chile}
\author[0000-0002-5991-8073]{Anibal Sierra}
\affiliation{Departamento de Astronom\'a, Universidad de Chile, Camino El Observatorio 1515, Las Condes, Santiago, Chile}
\author[0000-0003-2253-2270]{Sean M. Andrews}
\affiliation{Center for Astrophysics \textbar\, Harvard \& Smithsonian, 60 Garden St., Cambridge, MA 02138, USA}
\author[0000-0003-2251-0602]{John Carpenter}
\affiliation{Joint ALMA Observatory, Avenida Alonso de Córdova 3107, Vitacura, Santiago, Chile}
\author[0000-0002-1483-8811]{Ian Czekala}
\affiliation{Department of Astronomy and Astrophysics, 525 Davey Laboratory, The Pennsylvania State University, University Park, PA 16802, USA}
\affiliation{Center for Exoplanets and Habitable Worlds, 525 Davey Laboratory, The Pennsylvania State University, University Park, PA 16802, USA}
\affiliation{Center for Astrostatistics, 525 Davey Laboratory, The Pennsylvania State University, University Park, PA 16802, USA}
\affiliation{Institute for Computational \& Data Sciences, The Pennsylvania State University, University Park, PA 16802, USA}
\author[0000-0002-3393-2459]{Carsten Dominik}
\affiliation{Anton Pannekoek Institute for Astronomy, University of Amsterdam, Science Park 904, 1098XH Amsterdam, The Netherlands}
\author[0000-0002-1493-300X]{Thomas Henning}
\affiliation{Max Planck Institute for Astronomy, K\"onigstuhl 17, 69117, Heidelberg, Germany}
\author[0000-0002-1637-7393]{Francois Menard}
\affiliation{Univ. Grenoble Alpes, CNRS, IPAG, 38000 Grenoble, France}
\author[0000-0001-8764-1780]{Paola Pinilla}
\affiliation{Max Planck Institute for Astronomy, K\"onigstuhl 17, 69117, Heidelberg, Germany}
\affiliation{Mullard Space Science Laboratory, University College London, Holmbury St Mary, Dorking, Surrey RH5 6NT, UK}
\author[0000-0002-5903-8316]{Alice Zurlo}
\affiliation{N\'ucleo de Astronom\'a, Facultad de Ingenier\'a y Ciencias, Universidad Diego Portales, Av. Ejercito 441, Santiago, Chile}
\affiliation{Escuela de Ingenier\'a Industrial, Facultad de Ingenier\'a y Ciencias, Universidad Diego Portales, Av. Ejercito 441, Santiago, Chile}

\begin{abstract}
PDS\,70 is a unique system in which two protoplanets, PDS\,70\,b and c, have been discovered within the dust-depleted cavity of their disk, at $\sim$22 and 34\,au respectively, by direct imaging at infrared wavelengths. Subsequent detection of the planets in the H$\alpha$ line indicates that they are still accreting material through circumplanetary disks. In this Letter, we present new Atacama Large Millimeter/submillimeter Array (ALMA) observations of the dust continuum emission at 855\,$\mu$m at high angular resolution ($\sim$20\,mas, 2.3\,au) that aim to resolve the circumplanetary disks and constrain their dust masses. Our observations confirm the presence of a compact source of emission co-located with PDS\,70\,c, spatially separated from the circumstellar disk and less extended than $\sim$1.2\,au in radius,  a value close to the expected truncation radius  of the cicumplanetary disk at a third of the Hill radius. The emission around PDS\,70\,c has a peak intensity of $\sim$86$\pm$16\,$\mu \mathrm{Jy}~\mathrm{beam}^{-1}$ which corresponds to a dust mass of $\sim$0.031\,M$_{\oplus}$ or $\sim$0.007\,M$_{\oplus}$, assuming that it is only constituted of 1\,$\mu$m or 1 mm sized grains, respectively. We also detect extended, low surface brightness continuum emission within the cavity near PDS\,70\,b. We observe an optically thin inner disk within 18\,au of the star with an emission that could result from small micron-sized grains transported from the outer disk through the orbits of b and c. In addition, we find that the outer disk resolves into a narrow and bright ring with a faint inner shoulder.
\end{abstract}


%%%%%%%%%%%%%%%%%%%%%%%%%%%%%%%%%%%%%%%%%%%%%%%%%%%%%%%%%%%%%%%%%%%%%%%%%%%%%%%%

\section{Introduction}
Recent surveys have revealed that almost ubiquitously, protoplanetary disks appear highly structured with rings and gaps, spiral arms and asymmetries  \citep[e.g.,][]{garufi2018, andrews2020}. While other scenarios are discussed, these features are often interpreted as resulting from the presence of planets embedded in disks \citep[e.g.,][]{dong2015a, bae2018, Lodato2019}. Additional observational support for such a scenario can be found in the form of local perturbation of the gas velocity field from Keplerian rotation \citep{pinte2018, teague2019,casassus2019}. The quest to detect protoplanets embedded in their host disk through direct imaging has been challenging, with current detection limits on the order of a few Jupiter masses (M$_{\rm Jup}$) at large radii \citep[e.g.,][]{nuria2018, ruben2021}. A few protoplanet candidates have been claimed in the infrared (IR) and in the H${_{\alpha}}$ line \citep[e.g.,][]{sallum2015, reggiani2018} but remain controversial \citep{mendigutia2018}. 

The first robust detection through direct imaging techniques of a protoplanet still embedded in its natal disk  was obtained in the young system PDS\,70 \citep[spectral type K7; M$\sim$0.8\,M$_{\odot}$; age$\sim$5.4\,Myr old;][]{mueller2018} located at $\sim$112.4\,pc \citep{gaiaDR3} in the Upper Centaurus Lupus association \citep{Pecaut2016}. PDS\,70\,b was discovered with an orbital radius of $\sim$22\,au, and imaged at multiple IR wavelengths \citep{Keppler2018, mueller2018} as well as in a filter centered on the H$\alpha$ line \citep{Wagner2018b}. Subsequently, PDS\,70\,c was  discovered in H$\alpha$ imaging at the outer edge of the cavity with an orbital radius of $\sim$34\,au \citep{haffert2019}. These two planets carve a large cavity in the disk, evidenced by a cavity in dust \citep[e.g.,][]{hashimoto2012,Dong2012} and a gap in the $^{12}$CO gas emission along the orbit of PDS\,70\,b \citep{Keppler2019} that indicates significant gas depletion. Observations and hydrodynamic simulations indicate that the planets' orbital configuration is stable, close to a 2:1 mean motion resonance, with PDS\,70\,b in a slightly eccentric orbit  \citep[e$\sim$0.2;][]{bae2019, Toci2020, wang2021}. The masses of the two planets are still uncertain, although both planets are likely lighter than 10 M$_{\rm Jup}$ to ensure dynamical stability \citep{wang2021} and a non-eccentric outer disk \citep{bae2019}. Spectro-photometric analyses, limited to the IR regime (1-5\,$\mu$m) remain inconclusive, but suggest planet masses between 1 and a few M$_{\rm Jup}$ \citep[e.g.,][]{mueller2018,mesa2019,stolker2020} as well as a clear contribution from dust grains in clouds and/or circumplanetary disks (CPDs)  \citep{christiaens2019,stolker2020, Wang2020}.

\begin{table*}[t]
\centering
\caption{Summary of available ALMA Band 7 observations of PDS\,70. MRS is the maximum recoverable scale.}
\begin{tabular}{ccccccl}
\toprule
Label & ID & Date & Baselines & Frequency & MRS & References \\ 
 &  &  & [m]  &  [GHz] & [arcsec] &  \\ 
\hline
SB16 & 2015.1.00888.S & 2016 Aug 14-18 & 15-1462 &344-355 & 3.23 & Long et al. 2018\\ %(SB1) 
IB17  & 2017.A.00006.S & 2017 Dec 2-6 & 15-6855 & 346-357 & 1.05 & Keppler et al. 2019; Isella et al. 2019 \\ %(LB3)
LB19 & 2018.A.00030.S & 2019 Jul 27-31 & 92-8547 & 346-355 & 0.53 & This work\\\hline %(LB1) 
\end{tabular}
\label{tab:obsdata}
\end{table*}

CPDs play a fundamental role in planet formation, as they regulate the gas accretion onto the planet and determine the conditions for satellite formation. As gas enters the planet's sphere of influence, it falls at supersonic velocities onto the surface of the CPD \citep{tanigawa2012,judit2017}, possibly episodically \citep{Gressel2013}, leading to shocks that can ionize hydrogen and be traced in the H$\alpha$ line. From observations of the H$\alpha$ line, PDS\,70\,b and PDS\,70\,c are found to be accreting material from their host disk at a rate of $\sim10^{-8}$ M$_{\rm{Jup}}$ per year \citep{wagner2018, haffert2019, Thanathibodee2019, aoyama2019, hashimoto2020}. Using Atacama Large Millimeter/
submillimeter Array (ALMA) observations at $\sim$67\,mas$\times$50\,mas resolution, \citet{isella2019} showed evidence for sub-millimeter continuum emission co-located with PDS~70~c, interpreted as tracing a dusty CPD, and for another compact continuum emission source located at $\sim$74\,mas offset in a South West direction from b. The emission around c however was not spatially separated from the outer ring. In this Letter, we present new ALMA observations with 20\,mas resolution that provide an independent detection of a compact source of emission colocated with PDS\,70\,c and of  low surface brightness emission within the cavity close to PDS\,70\,b. The Letter is organized as follows.  Section~\ref{sec:obs} presents the observations and the procedure to calibrate the data. Section~\ref{sec:results} presents our new images and analysis. Finally, we discuss our findings in Section~\ref{sec:discussion}.

%%%%%%%%%%%%%%%%%%%%%%%%%%%%%%%%%%%%%%%%%%%%%%%%%%%%%%%%%%%%%%%%%%%%%%%%%%%%%%%%

\section{Observations}
\label{sec:obs}
This Letter presents new ALMA observations, hereafter referred to as LB19 (for 'Long Baselines 2019'), obtained in Band 7 ($\lambda=855\,\micron$), under a Director's Discretionary Time (DDT) program with ID 2018.A.00030.S. PDS\,70 was observed during 4 execution blocks (EB) with the C-8 configuration on 2019 July 27, 28, and 30, for a total on-source time of 43 minutes per execution. An observing log including the precipitable water vapor (PWV) levels and calibrator names is given in Appendix \ref{sec:log}. The spectral set-up was tuned to optimize continuum detection, but includes the $^{12}$CO J=3-2 line at 345.8\,GHz and the HCO$^+$ J=4-3 line at 356.7\,GHz, which  will be presented in forthcoming papers. The raw data calibration was done with the \texttt{CASA v.5.6.1} pipeline \citep{McMullin2007} and the self-calibration and post-processing imaging were done using \texttt{CASA v.5.4.0}. We first flagged the channels that included the $^{12}$CO and the HCO$^+$ lines and spectrally averaged the remaining channels to produce a continuum dataset. We imaged the resulting visibilities with the \texttt{tclean} task using the multi-scale \texttt{CLEAN} algorithm with scales of 0, 1, 3 and 6 times the beam FWHM, and an elliptic \texttt{CLEAN} mask encompassing the disk emission. To reduce the size of the data, we time averaged it to 6.06 seconds, i.e., 3 times the original integration time. After imaging, one EB image appeared of much lower SNR and therefore the corresponding visibilities were  rejected. The individual images of the three remaining execution blocks (EBs 0,1,3) did not appear astrometrically offset with respect to each other, which is as expected because they were taken very close in time. As the fluxes of all EBs match within 2\%, we concatenated the three EBs and self-calibrated them all together. To determine a good initial model for the self-calibration, we used multi-scale cleaning with the \texttt{tclean} task using a threshold of $\sim$7 times the rms noise level of the image. Using the tasks \texttt{gaincal} and \texttt{applycal}, we corrected for phase offsets between spectral windows, and between polarizations considering a solution interval of the scan length (\texttt{solint=inf}). Another iteration of phase self-calibration was done with a solution interval of 30s. We reached an overall improvement in peak SNR of 34\% after self-calibrating the LB19 data.  
%proper motion of pds70 is ra= -29.66 mas/yr$\pm$0.07 mas/yr dec=-23.82 mas/yr$\pm$0.06 mas/yr

\begin{figure*}[t] 
\begin{center}
\includegraphics[width=1.0\textwidth]{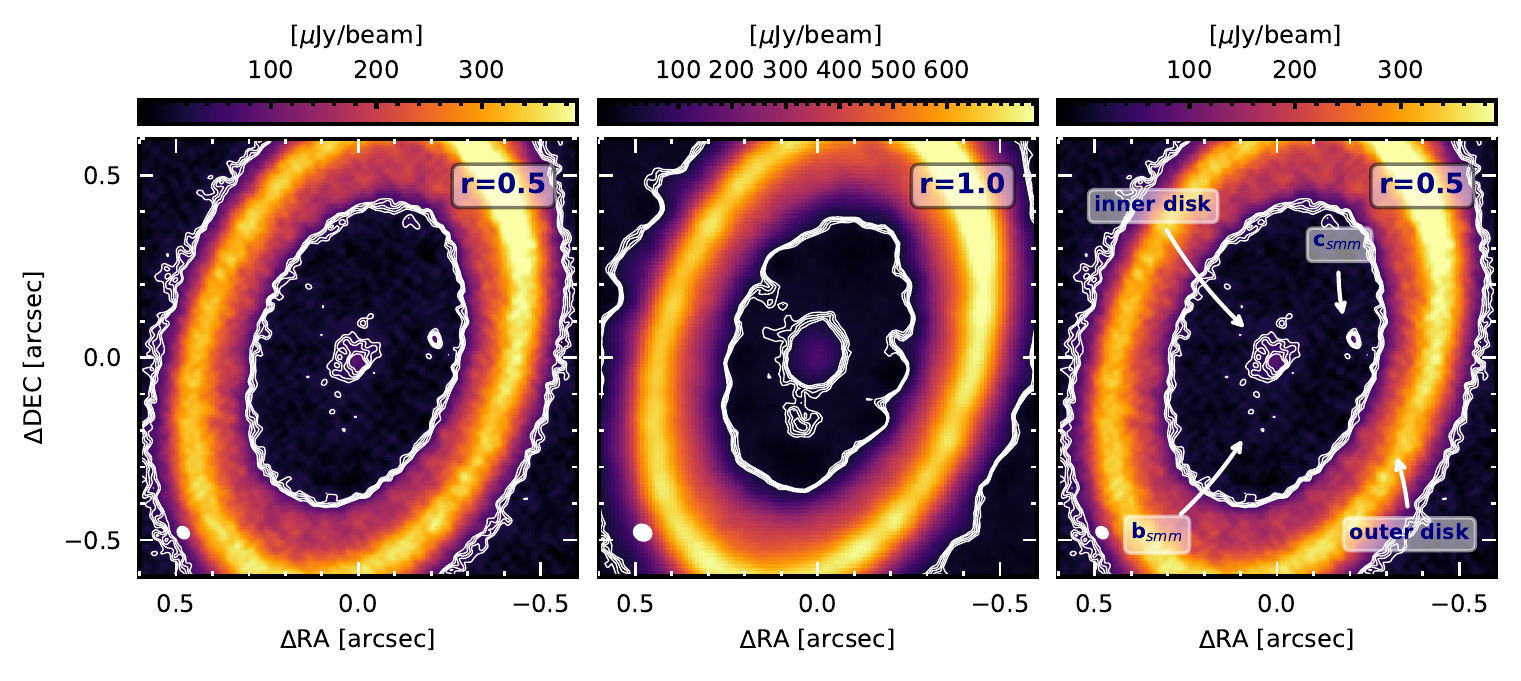}
\end{center}
\caption{Images of the new continuum observations of PDS\,70 (LB19+SB16). The data were imaged with a robust parameter of 0.5 (left) and 1 (center), with resolutions of 0.036\arcsec{}$\times$0.030\arcsec{} and 0.051\arcsec{}$\times$0.044\arcsec{}, respectively. The right panel shows the same image as in the left panel, with annotations. Beams are in the bottom left corner of each panel. Contours are 3 to 7$\sigma$, spaced by 1$\sigma$ (with $\sigma$=8.8 and 4.8\muJyb, respectively). An image gallery for all datasets is given in Appendix \ref{sec:Im}. }
\label{fig:continuum}
\end{figure*}

The LB19 data were combined with archival observations previously published in \cite{isella2019} and are summarized in Table\,\ref{tab:obsdata}. These observations correspond to program ID 2015.1.00888.S (PI: E. Akiyama), taken in August 2016 and labeled SB16 (for 'Short Baselines 2016'), and to program ID 2017.A.00006.S (PI: M. Keppler) taken in December 2017, labeled IB17 (for 'Intermediate Baselines 2017').  We refer the reader to Appendix A of \cite{isella2019} where the procedure for the self-calibration of SB16 and IB17 data is described in detail. For all datasets, we used the \texttt{statwt} task to weight the visibilities according to their scatter. Before combining the LB19 data with the previously published data, we fitted an elliptical ring to the maximum of the outer ring in the image plane, for all datasets separately, to derive the center of the image and then used the \texttt{fixvis} task to shift the image to the phase center, and assign it to a common phase center using the \texttt{fixplanets} task on the center coordinate derived by \cite{isella2019}. The fluxes of the executions in LB19 differed by $\sim$3\% from the archival datasets \citep[IB17+SB16; ][]{isella2019} and were rescaled using the DSHARP \texttt{rescale\_flux} function\footnote{\texttt{https://almascience.eso.org/almadata/lp/DSHARP/scripts/}}. After concatenation of the data, we followed the same procedure as explained above, with three rounds of phase self-calibration. %\ai{No amplitude selfcal?} \mb{It gave a very blobby ring. Would be worth trying out your new methodology.} 

We proceeded with imaging of the final data using \texttt{CLEAN}. In a normal \texttt{CLEAN} workflow, after the \texttt{CLEAN} iterations  terminate when the peak value of the residual image drops below a threshold value (4$\times$ rms noise level in the observations considered here), %\ai{We should \texttt{CLEAN} the final image much deeper. My guess is that if we \texttt{CLEAN} deep, the correction applied below becomes less important}, 
a restored \texttt{CLEAN} model is combined with the residual image to form the \texttt{CLEAN}ed image. As discussed in Czekala et al. \emph{in press}, however, the units of these two quantities differ: the units of the restored \texttt{CLEAN} model are $\mathrm{Jy}\,\{\mathrm{CLEAN\; beam}\}^{-1}$ while the units of the residual image are $\mathrm{Jy}\,\{\mathrm{dirty\;beam}\}^{-1}$, since it originated as the dirty image. When the \texttt{CLEAN} beam (typically chosen to be an elliptical Gaussian) poorly approximates the dirty beam (as is common with multi-configuration ALMA datasets), the normal \texttt{CLEAN} workflow produces a \texttt{CLEAN}ed image with an incorrect flux scale and compromised image fidelity, especially for faint emission. This phenomenon was first described in \citet{jorsater95}, and so we term it the ``JvM effect''. To correct for the unit mismatch, before combining the residual image with the restored \texttt{CLEAN} model, we first rescaled the residual image by the ratio of the \texttt{CLEAN} beam / dirty beam ``volumes'' (see ``JvM correction'', Czekala et al. \emph{in press}). 

To test the effect of the angular resolution on the image features and assess their robustness, we performed a grid of  \texttt{CLEAN}ed and JvM-corrected images, using Briggs weighting \citep{Briggs1992} with different robust parameters. A gallery of continuum images (and corresponding fluxes), synthesized from the new dataset alone (LB19) and from dataset combinations including the observations published by \citet{isella2019} (IB17+SB16; LB19+IB17+SB16) is given in Appendix \ref{sec:Im}.  Depending on the dataset and the robust parameter, our JvM-corrected images have a rms ranging between $\sim$4 and $\sim$26 $\mu  \mathrm{Jy}$ $\mathrm{beam}^{-1}$ across beam sizes of 93\,mas$\times$74\,mas to 20\,mas$\times$20\,mas (Table\,\ref{tab:imagingsummary}). We note that while the uv coverage and sensitivity are maximized when all datasets are combined (LB19+IB17+SB16), such a combination does not take into account the intrinsic changes of the emission that are due to the rotation of the system, and the change in the location of the dust surrounding the planets. Based on the orbital solutions of \citet{wang2021}, we expect a motion of $\sim$14\,mas for both planets between December 2017 and July 2019.

\begin{figure*}[t] 
\begin{center}
\includegraphics[width=0.7\textwidth]{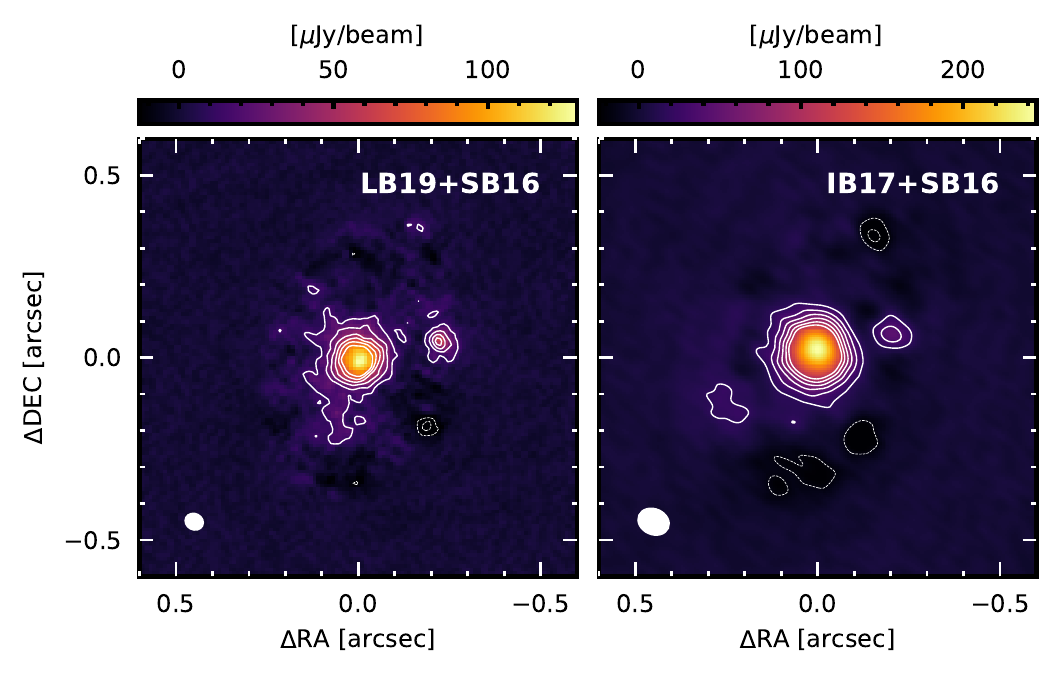}
\end{center}
%\caption{Residual images obtained after subtracting the Fourier transform of the \texttt{CLEAN} model for the outer ring (referred to as 'cavity images'), obtained with the new data (LB19+SB16; left) and the combined dataset  (LB19+IB17+SB16; right), with a Briggs robust parameter of 1. Contours are 3 to 18 times the rms noise level (4.7 and 4.3\muJyb, from left to right), spaced in steps of 3$\sigma$. Dashed contours correspond to -3$\sigma$. A gallery of cavity images is given in Appendix \ref{sec:Im}.}
\caption{Residual images obtained after subtracting the Fourier transform of the \texttt{CLEAN} model for the outer ring (referred to as 'cavity images'), obtained with the new data (LB19+SB16; left) and the data published in \citet{isella2019} (IB17+SB16; right) considering a Briggs robust parameter of 1. Contours are 3 to 18 times the rms noise level (4.7 and 6\,\muJyb, respectively), spaced in steps of 3$\sigma$. Dashed contours correspond to -3$\sigma$. A gallery of cavity images is given in Appendix \ref{sec:Im}.}
\label{fig:cavityJvMselection}
\end{figure*}

%This procedure, termed the ``JvM correction,'' substantially improved the image fidelity for our observations.

%%%%%%%%%%%%%%%%%%%%%%%%%%%%%%%%%%%%%%%%%%%%%%%%%%%%%%%%%%%%%%%%%%%%%%%%%%%%%%%%
%0.093\arcsec{}$\times$0.074\arcsec{} to 0.020\arcsec{}$\times$0.020\arcsec{}
\section{Results}
\label{sec:results}
\subsection{Continuum images}

Figure~\ref{fig:continuum} presents a selection of images of the continuum emission of PDS\,70 at 855\,$\mu$m, synthesized from the new ALMA observations combined with short baseline data (LB19+SB16). The disk is well detected with a spatially integrated flux density of $\sim$176$\pm$18 mJy (all images give similar values). After deprojecting the image with an inclination of $\sim$51.7$^\circ$ and a position angle of $\sim$160.4$^\circ$ \citep{Keppler2019}, we computed an azimuthally averaged radial profile and found that the outer disk resolves in a ring extending radially from $\sim$0.4\arcsec\ (45\,au) and $\sim$0.9\arcsec\ (100\,au).  The outer disk is not radially symmetric and shows a clear azimuthal asymmetric feature in the North West ($\sim$27\% brighter at peak compared to the mean ring value), as already discussed by \citet{long2018} and \citet{Keppler2019}. 
When imaged at high resolution, the outer disk resolves into a narrow and bright ring with a faint inner shoulder detected in the image at the 3-4$\sigma$ level  (Appendix \ref{sec:Im}). To better assess the presence of such substructures, we model the azimuthally averaged radial visibility profile using the \texttt{frank} package \citep{jennings2020}. Our analysis, presented in Appendix\,\ref{sec:frank} recovers a double peaked profile for the outer disk. Such a substructure was already hinted in the data presented in \citet{Keppler2019}. Inward of the outer disk, the dust-depleted cavity includes an inner disk that radially extends up to 0.16\arcsec{} (18\,au) and presents faint additional emission in the West and in the South of the inner disk that will be discussed in the next subsection.

\begin{figure*}[t] 
\begin{center}
\includegraphics[width=1.0\textwidth]{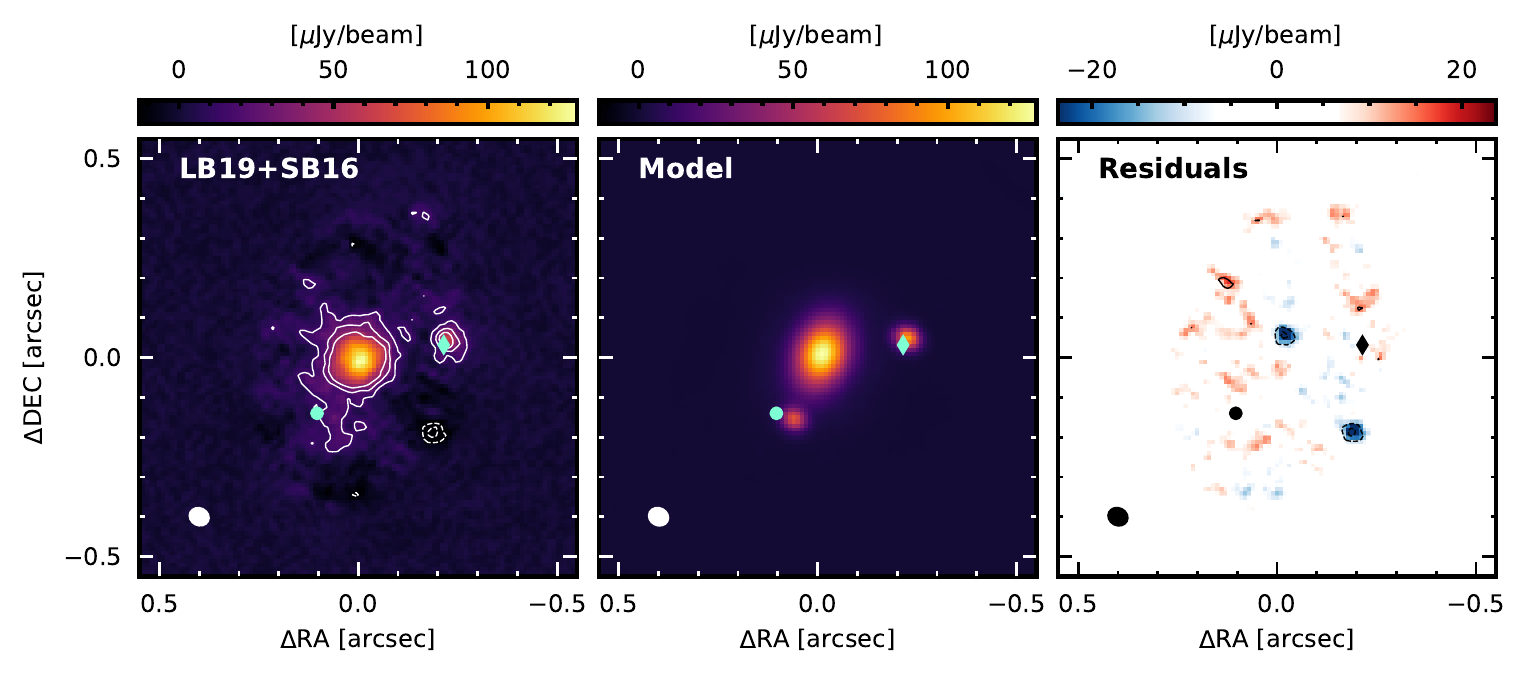}
\end{center}
\caption{From left to right: Cavity image for LB19+SB16; \texttt{Galario} best fit model for the inner disk, b$\rm_{smm}$, and c$\rm_{smm}$; Residuals from the \texttt{Galario} best fit model. All images are obtained with r=1. Contours are  3, 6, 9 $\sigma$. Dashed contours correspond to -6,-3$\sigma$.The predicted positions of the two planets in July 2019 are indicated with a circle and diamond (PDS\,70\,b and c, respectively).}
\label{fig:galario}
\end{figure*}

\subsection{Emission within the cavity}
Within the cavity, the inner disk appears well resolved with an integrated flux ranging between 727$\pm$27~$\mu$Jy and 888$\pm$59~$\mu$Jy  depending on the dataset (Table\,\ref{tab:innerdisk}).  When imaged at high angular resolution (e.g., Figure \,\ref{fig:continuum}, left), it appears irregular and the emission is discontinuous in the North.  

Continuum emission is also detected near the locations of the planets, confirming the findings of \citet{isella2019}. We use the same nomenclature as \citet{isella2019} and label the continuum emission located close to planet b and c, b$\rm_{smm}$ and c$\rm_{smm}$, respectively. 
The continuum emission around PDS\,70\,c, c$\rm_{smm}$, is  recovered in all images, and in particular in the standalone new, high resolution, dataset (LB19), where it appears as a 5.4 to 16$\sigma$ feature depending on the robust parameter. c$\rm_{smm}$ clearly separates from the outer disk when imaged at resolutions finer than $\sim$40\,mas. It appears unresolved even at our best angular resolution ($\sim$20\,mas;$\sim$2.3\,au). We find that its peak intensity is similar in all the images that spatially resolve it from the outer disk (see Appendix \ref{sec:cpdflux}), confirming its point-source nature. Depending on the dataset (IB17+SB16 or LB19+SB16) and the robust parameter, its peak intensity ranges between 80$\pm$6 and 107$\pm$15\,$\mu \mathrm{Jy}\mathrm{beam}^{-1}$. In the following, we will consider 86$\pm$16\,$\mu \mathrm{Jy}\mathrm{beam}^{-1}$ as a reference for further discussion. 

The emission located near PDS\,70\,b, b$\rm_{smm}$, is on the other hand, only recovered when the new high resolution data is combined with short baselines, and when the beam is larger than $\sim$50\,mas. This indicates that it is low surface brightness, extended emission. Its peak intensity and morphology vary greatly between images of different datasets (Table\,\ref{tab:innerdisk}), which makes its morphology and properties difficult to recover accurately. %, with an emission morphology and properties can not be accurately recovered in our images. %\mb{we'll need to explain why / how this is so. what does "sigma" mean in this context?  how is it measured?}

In order to assess whether the signal within the cavity could result from imaging artifacts, following \citet{andrews2018b}, we subtracted the Fourier transform of the \texttt{CLEAN} model of the outer disk, after blanking out the pixels within the cavity (using an elliptical mask of 0.25\arcsec{}$\times$ 0.4\arcsec{}), and image and model the visibilities carrying the residual signal from within the cavity. Figure~\ref{fig:cavityJvMselection} show two residual images, hereafter called 'cavity images', for LB19+SB16 and IB17+SB16, that clearly show that the inner disk emission and c$_{\rm smm}$ are recovered in both epochs, the latter with a significance up to 18$\sigma$. On the other hand, b$_{\rm smm}$ is detected at a 3$\sigma$ level only in some cavity images obtained from combined datasets. A gallery of cavity images are given in the Appendix \ref{sec:Im}.

\begin{table*}[t]
\centering
\caption{Best-fit parameters for the model to the cavity data for the datasets LB19+SB16 and IB17+SB16, with the $1\sigma$ error. The flux, radial peak position, and  width of the Gaussian for the inner disk are  $f_{\text{inn}}$, $r_{\text{inn}}$, $\sigma_{\text{inn}}$, respectively. The total flux and polar coordinates in the disk plane of b$\rm_{smm}$ and c$\rm_{smm}$ are $f_b$, $r_b$, $\theta_b$ and $f_c$, $\sigma_c$, $r_c$, $\theta_c$, respectively. The relative apparent astrometry $\Delta (\text{RA, Dec})$ is also provided.}
\begin{tabular}{c||ccc|ccc|ccc}
\hline
Dataset & $f_{\text{inn}}$ & $r_{\text{inn}}$ & $\sigma_{\text{inn}}$ & $r_b$ & $\theta_b$ & $f_b$     & $r_c$ & $\theta_c$ & $f_c$ \\ 
      & [$m$Jy]    & [mas]            & [mas]                 & [mas] & [deg]      & [$\mu$Jy] & [mas] &  [deg]     & [$\mu$Jy]  \\
\hline
%LB19+SB16 & $1.48_{-0.27}^{+0.03}$ & $2.0_{-1.6}^{+25.0}$ & $59.3_{-12.0}^{+2.7}$  & $178.5_{-3.8}^{+2.7}$& $174.0_{-1.3}^{+1.4}$ & $83.1_{-15.8}^{+12.4}$ & $324.9_{-2.7}^{+2.7}$  & $-70.6_{-0.7}^{+0.6}$   & $111.5_{-13.6}^{+14.0}$ \\
%IB17+SB16 & $1.66_{-0.32}^{+0.03}$ & $2.8_{-1.0}^{+24.8}$ & $52.6_{-12.0}^{+0.1}$  & --- & --- & --- & $329.4_{-10.1}^{+10.8}$ & $-68.9_{-1.0}^{+1.1}$   & $91.6_{-13.1}^{+14.4}$ \\
LB19+SB16 & $0.846_{-0.047}^{+0.036}$ & $2.0_{-1.6}^{+25.0}$ & $59.3_{-12.0}^{+2.7}$  & $178.5_{-3.8}^{+2.7}$& $174.0_{-1.3}^{+1.4}$ & $83.1_{-15.8}^{+12.4}$ & $324.9_{-2.7}^{+2.7}$  & $-70.6_{-0.7}^{+0.6}$   & $111.5_{-13.6}^{+14.0}$ \\
IB17+SB16 & $0.765_{-0.040}^{+0.018}$ & $2.8_{-1.0}^{+24.8}$ & $52.6_{-12.0}^{+0.1}$  & --- & --- & --- & $329.4_{-10.1}^{+10.8}$ & $-68.9_{-1.0}^{+1.1}$   & $91.6_{-13.1}^{+14.4}$ \\
\hline
     &  &  &  & $\Delta$\,RA [mas] & $\Delta$\,Dec [mas] &  & $\Delta$\,RA & $\Delta$\,Dec &  \\ 
\hline
LB19+SB16 &  &  &  & $70.1_{-2.5}^{+2.4}$ & $-163.0_{-3.2}^{+3.4}$ &  & $-215.1_{-1.6}^{+1.8}$ & $37.8_{-3.7}^{+3.3}$ &  \\ 
IB17+SB16 &  &  &  & ---                   & ---                    &  & $-219.2_{-6.5}^{+7.0}$ & $47.9_{-4.8}^{+4.9}$ &  \\ 
\end{tabular}
\label{tab:galario_results}
\end{table*}

As an additional test, we perform a model fit of the cavity visibilities using the dataset LB19+SB16, obtained after subtracting the Fourier transform of the \texttt{CLEAN} model of the outer disk using a robust parameter of 1. We consider a simple model for all three sources of emission within the cavity, namely the inner disk, b$\rm_{smm}$ and c$\rm_{smm}$, compute the Fourier transform using \texttt{galario} \citep{galario} and explore the parameter space using the Monte Carlo Markov chains implementation in \texttt{emcee} \citep{emcee}. Our model consists in a Gaussian ring for the inner disk, that enables to model an additional structure within the inner disk, a point source for c$\rm_{smm}$ (between PA=250$^\circ$ and 280$^\circ$), and a circular Gaussian for b$\rm_{smm}$, located in the South (between PA=70$^\circ$ and 250$^\circ$). A uniform prior was used over the allowed range for each parameter. Our best-fit model and residual maps are shown in Figure~\ref{fig:galario}, and corresponding parameters are in Table\,\ref{tab:galario_results}. We find that the best-fit location of c$\rm_{smm}$ is $\Delta (\text{RA, Dec})$=($-215.1_{-1.6}^{+1.8}$, $37.8_{-3.7}^{+3.3}$)~mas, close to the predicted position of PDS\,70\,c $\Delta (\text{RA, Dec})=(-214.8, 31.9)$~mas  (see Appendix \ref{sec:astrometry}). For b$\rm_{smm}$, the location is constrained to $\Delta (\text{RA, Dec})$=( $70.1_{-2.5}^{+2.4}$, $-163.0_{-3.2}^{+3.4}$)~mas, offset from the predicted position of PDS\,70\,b ($\Delta (\text{RA, Dec})=(96.9, -153.7)$~mas).

From the orbital fits of \citet{wang2021}, the expected motions of the planets between the epoch of the long baselines observations (December 2017 and July 2019) is similar for both, $\sim$14\,mas,  smaller than the angular resolution of our observations. To search for possible motion of c$\rm_{smm}$ between the two epochs, we performed the same modeling as above on the IB17+SB16 dataset. b$\rm_{smm}$ was not recovered in this fit, but the inner disk and c$\rm_{smm}$ were. Using the best-fit positions for c$\rm_{smm}$ at the two epochs, and considering a 2\,mas error in the centering of the two datasets, we find marginal evidence for a movement of the peak position of 10.9$\pm$6.9\,mas. We note that the nominal positional accuracy is defined as beam$_{\rm{FWHM}}$/SNR/0.9 \citep[][ and ALMA Cycle 8 2021 Technical Handbook]{astrometry}, with 0.9 a factor to account for a nominal 10\% signal decorrelation. We consider two images in which c$\rm_{smm}$ is imaged at a decent SNR and separated from the outer disk, LB19+SB16 (r=0.5) and IB17+SB16 (r=-0.3). With corresponding SNR of 8.9$\sigma$ and 7.1$\sigma$ on the peak intensity of c$\rm_{smm}$ respectively, and a beam FWHM of 36 and 60\,mas, respectively, the positional accuracies are $\sim$4.5\,mas and 9.4\,mas, respectively, comparable to the uncertainty that we derived for the apparent displacement of c$\rm_{smm}$. Additional observations with ALMA in the coming years, providing a longer time baseline, are needed to confirm such a movement.

\section{Discussion}
\label{sec:discussion}

\paragraph{A circumplanetary disk around PDS\,70\,c}
\citet{isella2019} reported the detection of c$\rm_{smm}$ using $\sim$67\,mas resolution observations. We confirm this detection with higher angular resolution observations that enable us to separate the emission from the outer disk. Given that the location of c$\rm_{smm}$ is very close to the existing H$\alpha$ and NIR measurements of PDS\,70\,c \citep{isella2019}, and to the expected positions of PDS\,70\,c at the time of our observations (Figure~\ref{fig:galario}), we interpret it as tracing the millimeter emission of dust grains located in a CPD. Assuming that c$\rm_{smm}$ is optically thin, its flux density can be converted into a dust mass estimate, for a given dust opacity and temperature. We note that if the emission is optically thick, such an assumption would provide a lower limit in the dust mass. The CPD temperature is also uncertain. It is determined by the sum of various sources of heating, namely viscous heating due to accretion of material through the CPD, accretion shocks, and external irradiation from both the planet and the star \citep{isella2014, isella2019, andrews2021}. Using 2\,M$_{\rm{Jup}}$, 2\,R$_{\rm{Jup}}$, and 1055~K as the mass, radius, and temperature of PDS\,70\,c \citep{wang2021}, a mass accretion rate of 10$^{-8}$\,M$_{\rm{Jup}}$/year \citep{haffert2019}, we find that at a radial distance of 1 au from the planet, T$_{vis}$ = 3\,K, and T$_{p,irr}$= 18\,K. Considering a stellar-irradiation temperature of T$_{s,irr}$ = 24\,K at the location of PDS\,70\,c (obtained from the radiative transfer model of \citet{Keppler2019}), the  CPD temperature at 1\,au is T$_{CPD}^{4}$ = T$_{vis}^{4}$ + T$_{p,irr}^{4}$ + T$_{s,irr}^{4}$, that is T$_{CPD} \sim$ 26\,K. Considering a typical dust opacity for 1\,mm sized grains of 3.63 cm$^{2}$\,g$^{-1}$ \citep{Birnstiel2018} and a temperature of 26\,K, we estimate a CPD dust mass of $\sim$0.007\,$M_{\oplus}$. A lower dust mass would be inferred if the dust temperature is higher than considered here \citep{schulik2020}. 
%as accretion shocks due to free-falling gas can however locally heat the CPD to temperatures as high as a few hundreds of K \citep{schulik2020}.  
%, although it is unclear if viscous heating could dominate the temperature budget with the estimated mass accretion rates from H$\alpha$ line observations \citep{haffert2019, isella2019}. 

However, PDS\,70\,c is massive enough to carve a gap, and, as a consequence, large grains are trapped in a pressure maximum in the outer disk while small grains, well coupled to the gas, can flow inward. This is confirmed by the different cavity outer radii measured in scattered light compared to mm wavelengths \citep[probing small and large grains, respectively;][]{Keppler2019}. The CPD is therefore only replenished with small dust particles  that leak into the cavity \citep{bae2019} through meridional flows from the upper protoplanetary disk layers \citep[e.g.,][]{kley2001,ayliffe2009}. If the CPD contains only small 1\,$\mu$m sized grains \citep[with an opacity of 0.79 cm$^{2}$ g$^{-1}$;][]{Birnstiel2018} the CPD dust mass increases to $\sim$0.031\,M$_{\oplus}$. It is of course possible that the CPD hosts a range of particle sizes if the grains can grow. \citet{bae2019} find that, if a steady state is achieved between the mass inflow to the CPD and the mass accretion rate onto the planet, the amount of sub-micron grains in the CPD would largely underestimate the observed mm flux and that accumulation of grains beyond the steady-state amount and/or in-situ grain growth is needed to account for it. In Appendix \ref{sec:cpdmass}, we show the range of dust masses that the CPD would have for various dust grain size distributions, as a function of the maximum grain size.  With these mass estimates, the ratio between the CPD dust mass and the planet mass, considering 2\,M$_{\rm{Jup}}$ \citep{wang2021}, ranges between 1 and 5$\times$10$^{-5}$. %, close to what is found for Jupiter's moons. 

If small grains can grow to mm sizes within the CPD, they could rapidly be lost as they efficiently drift toward the planet and it only takes 100-1000 years for an accreting CPD to lose all its mm dust \citep{zhu2018}. However, as in protoplanetary disks, local gas pressure maxima can act as particle traps, and prevent these grains from drifting. Interestingly, this can naturally occur in CPDs. Most of the gas that is feeding the CPD through meridional flows is then radially flowing outward in a decretion disk. The balance between the sub-Keplerian headwind and viscous outflow associated with a decretion flow leads to a global dust trap \citep{batygin2020}. As a consequence, dust grains with sizes 0.1-10\,mm may be trapped in the CPD and as the dust-to-gas ratio increases, streaming instabilities might be triggered \citep{joanna2018}, or gravitational fragmentation in the outer regions of the CPD \citep{batygin2020} that will eventually lead to the formation of satellitesimals. At the same time, dust particles can accrete via pebble-accretion onto the satellitesimals formed in
situ or captured from the disk edge \citep[e.g.,][]{ronnet_2020}. %\mb{comment on metallicity? comment on OTS44 as another one of these disks?} 
%\ai{I wonder what dust-to-gas ratio  do you need for SI to occur and whether this is consistent with having a large dust trap in the outer disk that might prevent most of the dust  from accreting onto the CPD. Do current models take this into account?}

Our observations also put a strong constraint on the spatial extent of the CPD as seen in the dust emission at mm wavelengths. The emission c$\rm_{smm}$ is unresolved even at our highest angular resolution, and its peak intensity is similar over a range of beam sizes, until $\sim$40\,mas, beyond which the CPD does not separate from the outer disk anymore (Appendix \ref{sec:cpdflux}). This indicates that it is more compact than 1.2\,au in radius. On the other hand, there is a lower limit to the CPD extent needed to account for the observed flux. Assuming that it is a uniform disk with an optical depth of 1, and considering a temperature of 26\,K, we find that it has a radius of 0.58\,au. These two values (0.58 and 1.2\,au) are therefore the lower and upper limits on the CPD radial extent constrained from our observations. The CPD is expected to be truncated (in gas) at a third of the Hill radius, which for PDS\,70\,c, assuming a planet mass of 2\,M$_{\rm{Jup}}$ at 34\,au, is 1/3 $\times$3.1 $\sim$1\,au. 3D simulations show that isothermal CPD are bound within 10\% of the Bondi radius \citep{fung2019}, that is 1/10 $\times$11 $\sim$1.1\,au for PDS\,70\,c assuming a local temperature of 26~K. Both estimates are therefore consistent with our constraints. 
%If the CPD is optically thick, it would be more compact than 0.4\,au, and its extent would be $\sim0.4\times(26\,{\rm K}/T)^{1/2}\,$au, with $T$ its temperature. 
However, we cannot rule out the possibility that the gas component of the CPD extends beyond the dust component, in particular if some dust grains in the CPD drift inward.

% Figure~\ref{fig:overlayPM} shows an overlay of the cavity images for the data published in \citet{isella2019} (December 2017; IB17+SB16) and the new ALMA observations (July 2019; LB19+SB16), both combined with the short baseline data. The two images are consistent for both the inner disk emission as well as c$\rm_{smm}$, making the latter a robust detection. \todo{Move PM figure to main text.}

\begin{figure*}[t] 
\begin{center}
\includegraphics[width=0.95\textwidth]{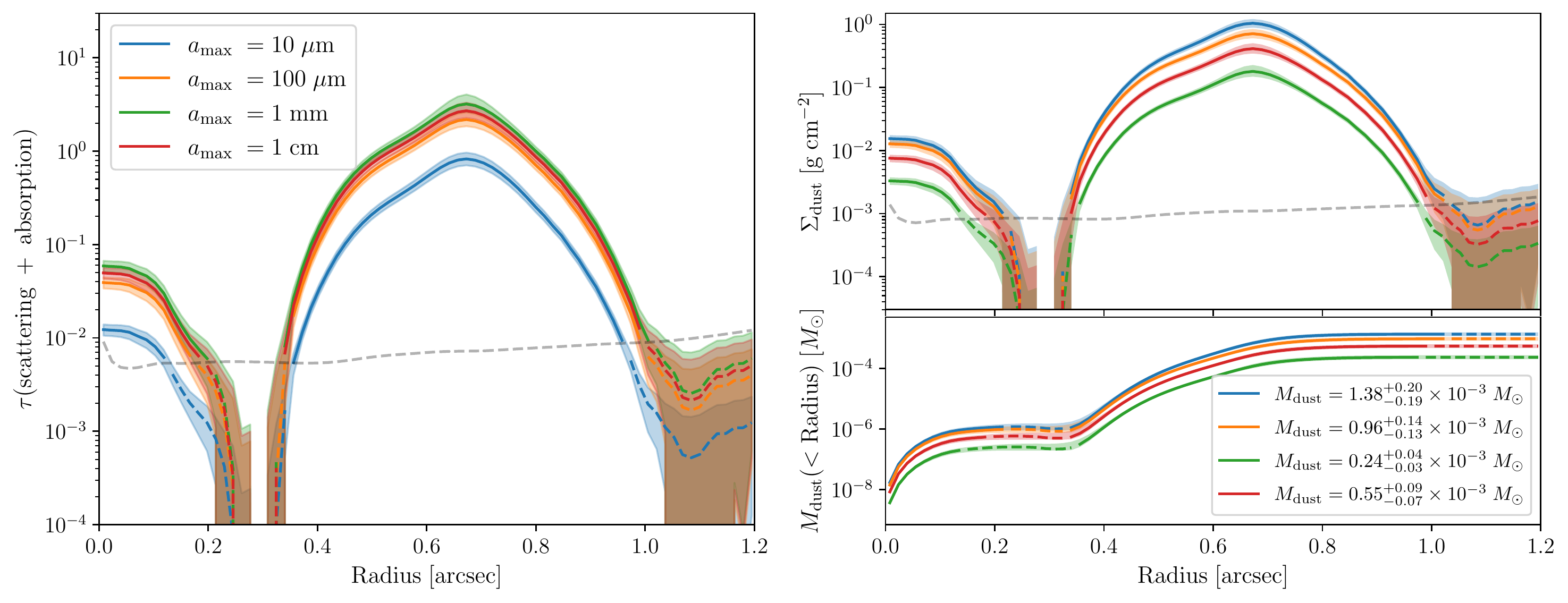}
\end{center}
\caption{Left: Total optical depth of the continuum emission computed from the azimuthally averaged radial profile of the r=1 image of LB19+IB17+SB16. The lines show 4 models with different maximum grain sizes. The grey dashed line corresponds to a floor value of 3$\sigma/\sqrt{N}$, with $\sigma$ the image rms, and N the number of beams in a radial bin. Shaded regions indicate error bars, computed as the square root of the quadratic sum of the image rms, the standard deviation in the radial bin and the 10\% flux uncertainty. Right: dust surface density profiles (top) and corresponding cumulative masses (bottom). }
\label{fig:TauCumMass}
\end{figure*}

\paragraph{Extended faint emission near PDS\,70\,b.}
The nature of the material close to PDS\,70\,b is unclear. It is not detected in the images obtained at high resolution with small synthesized beams but is apparent at low SNR at intermediate resolution indicating that it has a low surface brightness. It is confirmed in the two epochs 2017 and 2019, when combined with the short baselines data.
b$\rm_{smm}$ appears offset toward the South-West from the position of PDS\,70\,b, confirming the findings of \citet{isella2019} who speculated that it could be tracing dust trapped at the Lagrangian point L5  \citep{Montesinos2020}, if the planet is on an inclined orbit.  
The shape of the b$\rm_{smm}$ in our images is suggestive that it could also trace the faint signature of a streamer connecting the planets to the inner disk. Evidence for dust grains in the vicinity of PDS\,70\,b is clear already from the IR spectral energy distribution \citep{stolker2020, wang2021}, likely explaining the non-detection of Br$\gamma$ \citep{christiaens2019} and H$\beta$ emission lines \citep{hashimoto2020}. It is interesting to understand why PDS\,70\,b, at the sensitivity of our observations, does not seem to host a compact, dusty, circumplanetary disk as PDS\,70\,c does. A possibility would be that PDS\,70\,b has a much smaller Hill radius than PDS\,70\,c, as it orbits at smaller separation.  %A possibility would be that PDS\,70\,b is of lower mass than PDS\,70\,c, thus implying a smaller Hill radius. However, this seems at odds with the ALMA CO integrated intensity maps, in which a deep gap co-located with the orbital radius of planet b is clearly detected, suggesting on the contrary that  PDS\,70\,b is more massive than PDS\,70\,c \citep{Keppler2019, Facchini2021}. 
Another natural explanation could be that PDS\,70\,b is starved of dust grains, as only the small grains that leak through the orbit of PDS\,70\,c and are transported through a streamer from the outer to the inner planet would enter the region of influence of PDS\,70\,b. Finally, it could be that the nature of the CPD is different around the two planets, with a decretion disk around PDS\,70\,c, and an accretion disk around PDS\,70\,b that is fed through a streamer coming from PDS\,70\,c rather than through meridional flows. More theoretical work looking at formation of CPDs in systems hosting two giant planets is needed to assess the potential differences between CPD formation in the inner and outer planet.
%\todo{Cite Shibaike2017; Seba's paper on CPD}

\paragraph{Inner disk.} An inner dusty disk, evidenced in the IR spectral energy distribution and scattered light images is also clearly detected in our images up to $\sim$0.16\arcsec{} ($\sim$18\,au) \citep[see also,][]{long2018, Keppler2019}. Considering that the planets are filtering material from the outer disk such that only small dust particles can flow in the cavity, as for the CPD, it is unclear whether the inner disk mm emission is due to a population of small or large dust grains. To address this question, we computed the dust surface density and optical depth radial profiles of the continuum emission, using the combined dataset (SB16+IB17+LB19) imaged with robust=1. We consider 4 models for the dust grain population, that follow a size distribution $n(a)da \propto a^{-3.5}da$ with a maximum grain size $a_{\rm{max}}$ of 10\,$\mu$m, 100\,$\mu$m, 1\,mm and 1\,cm, and a minimum size of 0.05\,$\mu$m. We use the DSHARP opacities \citep{Birnstiel2018} and the temperature profile output of the radiative transfer model of \citet{Keppler2018}. % that fits the spectral energy distribution of PDS\,70 and the infrared scattered light images. %The albedo and opacity coefficient are determined by the maximum grain size, which is fixed in the disk. 
The dust surface density as well as the total optical depth $\tau_{\nu}$ is numerically computed, considering scattering and  absorption opacities \citep{Sierra2020, sierra2021}. Figure~\ref{fig:TauCumMass}, left, shows the total optical depth $\tau_{\nu}$ for all 4 models. The right panels show the dust surface density profiles (top) and corresponding cumulative masses (bottom). The dust surface density is maximum at the outer disk that is obviously the disk region that contributes to most of the dust mass ($\sim$0.24$\times$10$^{-3}$\,$M_{\odot}$ for $a_{\rm{max}}$= 1\,mm). We note that without the inclusion of scattering, the optical depth would follow the curve of the dust population with $a_{\rm{max}}$=10\,$\mu$m, as the albedo at mm wavelengths is negligible for these small grains. In all these models, the inner disk is optically thin, with a total dust mass of $\sim 2\times10^{-7}-10^{-6}$\,$M_{\odot}$ (i.e., 0.08-0.36$M_{\oplus}$). 

It therefore appears that the emission at 855\,$\mu$m from the inner disk regions located within the orbit of PDS70\,b could be accounted for by a population of small grains. Interestingly, we note that the near infrared excess apparent in the spectral energy distribution of PDS70 is very low \citep{Dong2012}. This emission is mostly due to the thermal emission of small grains located within the innermost au and such a low excess could indicate a low small-dust mass content in the inner disk, and therefore suggest the additional presence of larger dust grains in order to account for the measured flux at 855\,$\mu$m. However, the inner disk emission in the infrared could still be optically thick \citep{Dong2012}, making it difficult to directly relate to our sub-millimeter observations and multiple wavelength observations in the millimeter regime are needed to constrain the grain size population in the inner disk. %Multiple wavelength observations are needed to constrain the grain size population, it appears that the emission from the inner disk can be well accounted for by a population of small grains. 
We note that the brightness temperature might be underestimated near the star because of our limited angular resolution and that it is possible that the innermost disk regions are optically thick also at sub-millimeter wavelengths. 
%In a generic study to explain the presence of inner disks in the large cavities of transition disks, \citet{pinilla2016} showed that within the snow line, dust grains fragment at lower velocities, allowing small particles to remain in the inner disk for a few millions of years. 
The longevity of the inner disk remains unclear; the replenishment flow is controlled by the planets, if it is so strongly depleted (in gas) it may not allow grains to grow efficiently. It is possible that some of the dust in the inner disk is of second generation produced by collision of larger bodies, perhaps stirred up by PDS\,70\,b. The star exhibits a small, but non negligible, mass accretion rate, for which an additional mass reservoir in the inner disk, such as a dead zone, was recently suggested \citep{thanathibodee2020}. Determining the physical conditions there-in, in particular the dust to gas ratio, would be crucial for understanding whether such an inner disk can still grow terrestrial planets within a system hosting two outer giant planets. The current dust mass estimates are so low that it is unlikely that planets could form through pebble accretion \citep{Lambrechts2019}.  
% According to your dust mass calculations of the inner disk, you won't have enough mass in dust to form planets (no even Mars) through pebble accretion (Lambrechts +2019)

\paragraph{Outer disk structure.} Our observations at high angular resolution indicate that the outer disk hosts substructures. In addition to an 'arc' in the North-West, already seen at lower resolution images \citep{long2018, Keppler2019}, it resolves into two components, that can be either a double-ring structure with a dip at $\sim$0.55\arcsec{} or a bright ring with an inner shoulder. %Such a two-component continuum emission is also seen in other transition disks although with an outer shoulder \citep{pinilla2018a, Huang2020}. 
Interestingly, \citet{Huang2020} also find with high resolution observations, a two-component structure in GM\,Aur, with a bright ring and an outer shoulder. It is unclear if such  two-component structure in PDS\,70 could be due to a secondary gap induced by PDS\,70\,c as an outer secondary gap opens only when the disk is sufficiently cold \citep{bae2018b}, with $(h/r)_p \lesssim 0.06$ where $(h/r)_p$ is the disk aspect ratio at the location of the planet \citep[$(h/r)_p \simeq 0.08$ at PDS\,70\,c's location;][]{bae2019}. On the other hand, recent three-dimensional planet-disk interaction simulations including both gas and dust components showed that dust grains at the gap edge can have radial structures \citep{bi2021}, potentially induced by  corrugated vertical flows driven by the spiral wave instability \citep{bae2016a,bae2016b} or meridional flows \citep{fung2016}. Alternatively, such a substructure could be due to the presence of an additional, yet-undetected low-mass planet embedded within the outer disk. Similar multiple-ring substructures were also observed in other transition disks, such as HD\,169142 in which three narrow rings were found and interpreted as tracing a migrating 10\,M$_{\oplus}$ in a low viscosity disk \citep{seba2019}. However, hydrodynamical simulations show that thermodynamics can dramatically affect the structure of gas and dust, with different disk cooling timescales leading to different planet-induced substructures \citep{facchini2020}. Further chemical surveys will help to constrain the density and temperature structures \citep{Facchini2021}, enabling to test the possibility that an additional, low-mass planet is responsible for the structured outer ring and constrain the mass and radial location of that planet.
We note that it is unlikely that an additional planet within the outer continuum ring disrupt the planetary system. In a two-planet system neglecting the eccentricity damping from the protoplanetary disk gas, the planets can avoid close encounters and are Hill-stable when their orbital separation is greater than $3.46~R_{\rm H}$, where $R_{\rm H} = a_1[(M_1 + M_2) / 3M_*]^{1/3}$ is the mutual Hill radius \citep{gladman1993,barnes2006}. The addition of a third planet generally makes the stability criteria more stringent because the conservation of the total angular momentum and energy can no longer guarantee the avoidance of close encounters even for initially large separations beyond the Hill-stability criteria \citep{tamayo2015}. However, provided that the protoplanetary disk gas provides sufficient eccentricity damping, \citet{tamayo2015} argued that the two-planet criteria can still be used in three-planet cases. Assuming a range of $1-10~M_{\rm Jup}$ for PDS\,70\,c and a Saturn mass for the hypothesized additional planet, this criteria is met when the latter is located beyond $44 - 53$~au. Therefore, the system would be dynamically stable if the additional planet is located within the dip in the outer continuum ring at $\sim 60$~au. Future numerical simulations will allow to further test our conclusions. 

\section{Conclusions}
In this Letter, we report new ALMA observations obtained at high angular resolution ($\sim$20\,mas) at 855\,$\mu$m of the PDS\,70 system. We confirm the tentative detection by \citet{isella2019} of a compact source co-located with the position of PDS\,70\,c with an independent dataset at higher angular resolution. These new observations provide the most compelling evidence of the presence of a CPD around an accreting planet to date. Future  molecular line infrared observations at very high angular resolution may be able to detect rotating gas around PDS\,70\,\,c, providing conclusive results on the nature of the continuum mm emission. The detection of unresolved ($r<1.2\,$au) emission around planet c confirms that circumplanetary material is able to retain dust for long timescales, as required in satellite formation models. 

These ALMA observations shed new light on the origin of the mm emission close to planet b. The emission is diffuse with a low surface brightness and is suggestive of a streamer of material connecting the planets to the inner disk,  providing insights into the transport of material through a cavity generated by two massive planets. The non-detection of a point source around PDS\,70\,b indicates a smaller and/or less massive CPD around planet b as compared to planet c, due to the filtering of dust grains by planet c preventing large amounts of dust to leak through the cavity, or that the nature of the two CPDs differ. We also detect a faint inner disk emission that could be reproduced with small 1\,$\mu$m dust grains, and resolve the outer disk into two substructures (a bright ring and an inner shoulder). % that might indicate the presence of a yet undetected additional planet within the outer disk.

PDS 70 is the best system to date to study and characterize circumplanetary disks, but also planet-disk interactions and disk cavity clearing by massive planets. The two massive planets, likely migrating outward in a grand tack-like scenario \citep{bae2019}, are reminiscent of the Jupiter-Saturn pair, at larger distances from the star. Detailed studies of the circumplanetary disks, and of the leakage of material through the cavity, will provide strong constraints on the formation of satellites around gas giants, and on the ability to provide the mass reservoir needed to form terrestrial planets in the inner regions of the disk. Upcoming studies of the gas kinematics and chemistry of PDS\,70 will complement the view provided by this work, serving as a benchmark for models of satellite formation, planet-disk interactions and delivery of chemically enriched material to planetary atmospheres.

% \mb{rewrite}
% In this Letter, we report on new ALMA observations obtained at high angular resolution ($\sim$20\,mas) at 855\,$\mu$m, and confirm the presence of a dusty circumplanetary disk around the planet PDS\,70\,c with a peak intensity of 86$\pm$16\,$\mu \mathrm{Jy}\mathrm{beam}^{-1}$. This CPD appears unresolved, and is more compact than 1.2\,au. Assuming optically thin emission from 1 mm sized grains, the CPD mass is $\sim$0.006M$_{\oplus}$, and the CPD to planet mass ratio is close to 10$^{-5}$. We also find extended faint emission close to PDS\,70\,b but its nature is unclear, and it could be tracing a faint streamer connecting the planets to the inner disk. The differences between the sub-mm emission around PDS\,70\,b and PDS\,70\,c might be due to the fact that PDS\,70\,c lies by the mass reservoir in the outer disk, while PDS\,70\,b can only be fed through the spiral arms from PDS\,70\,c with small particles that leak through the cavity. Another possible difference can lie in their ability to trap particles in their CPD. 
% We also detect a faint inner disk emission that can be reproduced with small 10\,$\mu$m dust grains, and resolve the outer disk into two substructures (a bright ring and an inner shoulder) that might indicate the presence of a yet undetected additional planet within the outer disk. \mb{Add a nice conclusion about complementary of kinematics, coming works.}

\appendix

\section{ALMA observations, images and fluxes}
\subsection{Observing log of the new ALMA observations}
\label{sec:log}
Table\,\ref{tab:obslog} provides the observing log of the new ALMA observations presented in this Letter. EB2, indicated in italic, was not included in the images.  

\begin{table*}[!h]
\centering
\caption{Summary of new continuum ALMA observations presented in this Letter, labeled LB19. EB2 was rejected. }
\begin{tabular}{ccccccc}
\toprule
Date	&    Antennas & Baselines [m] 	 & Time [min] & Mean PWV [mm]  & Bandpass/flux  & Phase calibrator \\
\hline
EB0: 27 July 2019 & 41 & 92-8283 & 43 & 0.6 & J1427-4206 & J1407-4302\\
%20:51-22:18 
EB1: 27 July 2019 & 41 & 92-8283 & 43 & 0.6 & J1427-4206 & J1407-4302\\
\textit{EB2: 28 July 2019} & 45 & 92-8547 & 43 & 0.4 & J1427-4206 & J1407-4302\\
EB3: 30 July 2019 & 43 & 92-8547 & 43 & 0.7& J1427-4206 & J1407-4302\\\hline
\end{tabular}
\label{tab:obslog}
\end{table*}

\subsection{Image galleries and corresponding fluxes}
\label{sec:Im}
\begin{figure*}[!h]
\begin{center}
\includegraphics[width=1.0\textwidth]{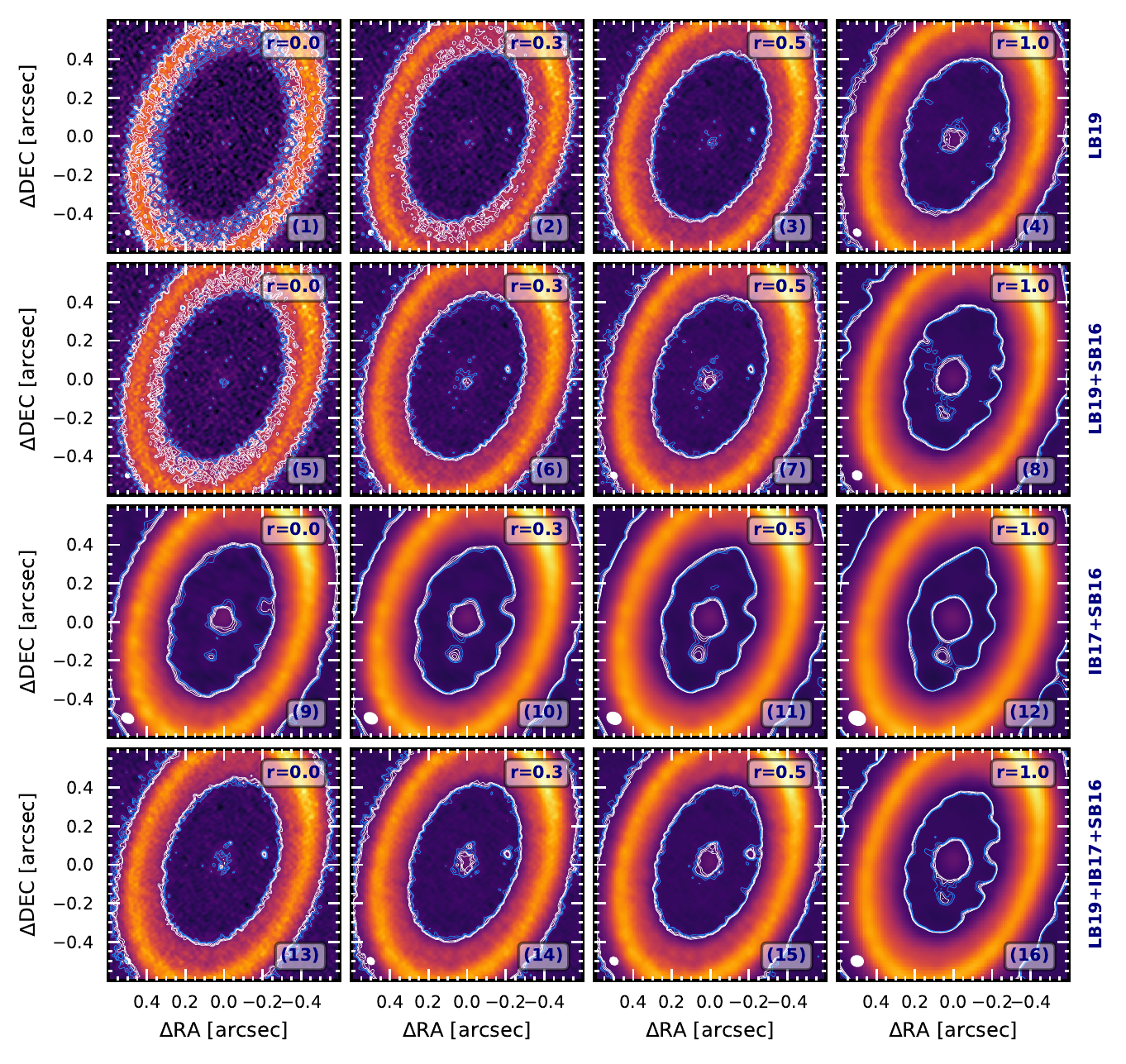}
\end{center}
\caption{Gallery of images for all datasets. 3 and 4$\sigma$ contours and  5, 6, 7$\sigma$ contours are showed in blue and white, respectively. Rows correspond to different datasets, while columns are for different Briggs robust values (from 0 to 1; from left to right).}
\label{fig:allJvM}
\end{figure*}

\begin{figure*}[t] 
\begin{center}
\includegraphics[width=0.8\textwidth]{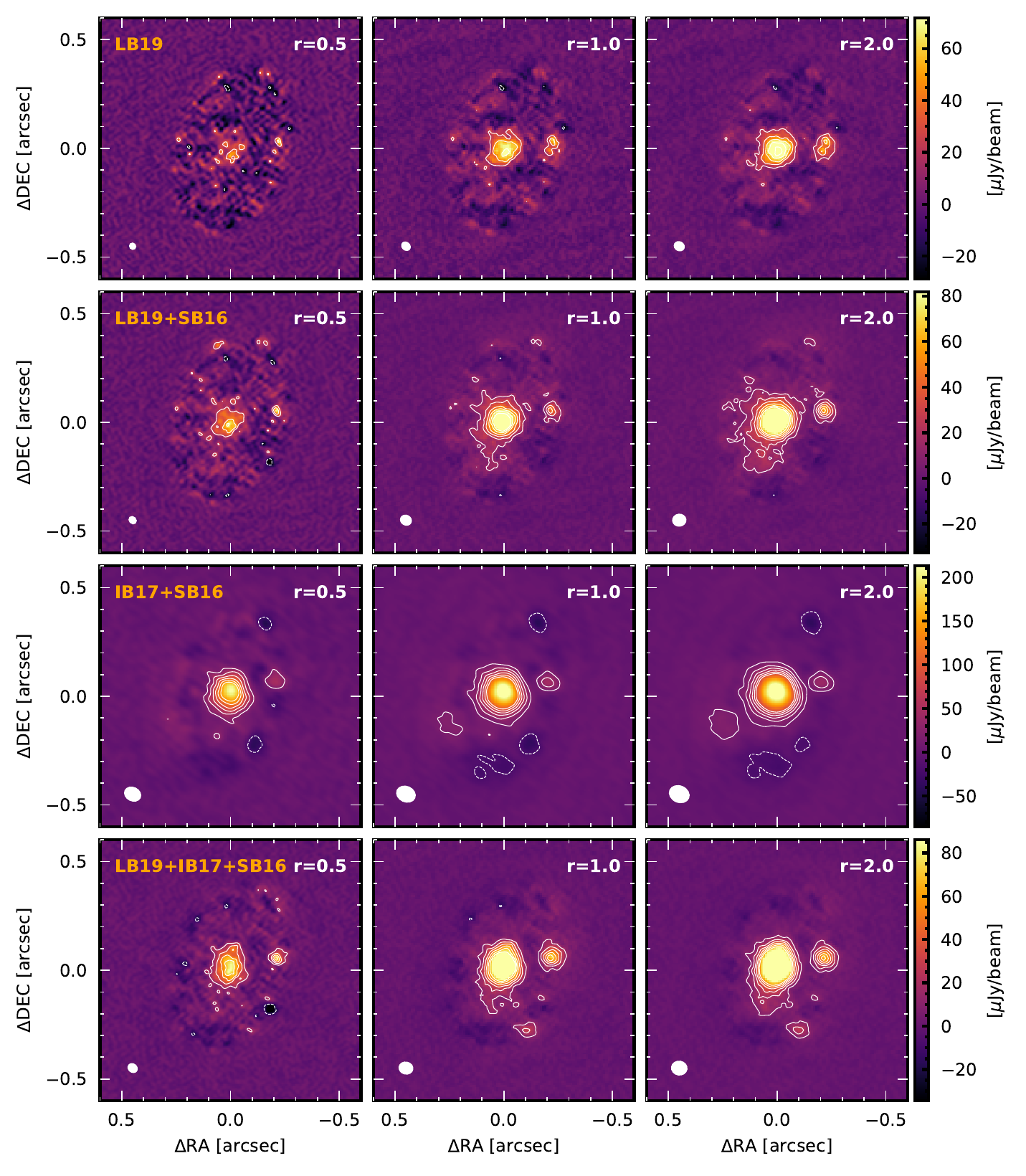}
\end{center}
\caption{Gallery of cavity images. Contours are 3 to 18$\sigma$, spaced by 3$\sigma$. Dotted lines traces contours at -3$\sigma$. Rows correspond to different datasets, while columns are for different Briggs robust values (0.5, 1, 2, from left to right).}
\label{fig:cavityJvM}
\end{figure*}

To test the effect of the angular resolution on the image features and assess whether they are recovered in various images, we performed a grid of \texttt{CLEAN}ed and JvM-corrected images using different datasets, and different Briggs robust parameters. Figure \ref{fig:allJvM} presents  the resulting images. Corresponding image properties and fluxes are reported in Tables\,\ref{tab:imagingsummary} and  \ref{tab:innerdisk}. Figure \ref{fig:cavityJvM} shows the residual images (called 'cavity images') obtained after subtracting the Fourier transform of the \texttt{CLEAN} model of the outer disk, for robust values of 0.5, 1 and 2.0.

\begin{table*}[!h]
\centering
\caption{Summary of disk and CPD properties for various datasets. The c$\rm_{smm}$  peak intensities were computed with the CASA task \texttt{imstat} in an aperture centered on the CPD, with major/minor axis twice the beam major/minor axis (col. 7) and with a Gaussian fit when possible (col. 8). The rms is computed considering an annulus between 2.4\arcsec{} and 6\arcsec{}. 
We considered 10\% calibration uncertainty as the flux uncertainty.} 
\begin{tabular}{cccc|cc|cc}
\hline \hline
\multicolumn{4}{c}{{\bf Obs.}} & \multicolumn{2}{c}{{\bf Disk}} & \multicolumn{2}{c}{{\bf Emission around PDS70\,c}} 	\\

Dataset & Briggs 	&  Beam, PA & rms noise &  Peak $I_{\nu}$ & Total Flux &  Peak $I_{\nu}$  & Gauss fit peak $I_{\nu}$   \\
& par. &  [mas $\times$ mas] & [$\mu  \mathrm{Jy}$ $\mathrm{beam}^{-1}$] & [m$\mathrm{Jy}$ $\mathrm{beam}^{-1}$ ] & [mJy] & [$\mu  \mathrm{Jy}$ $\mathrm{beam}^{-1}$] & [$\mu  \mathrm{Jy}$ $\mathrm{beam}^{-1}$]  \\
\hline
LB19 & 0	& 22$\times$22, 29$^\circ$ & 20.4 & 0.29 &  172$\pm$17 & 90$\pm$20 & 91$\pm$10 \\
     &0.3   & 26$\times$25, 31$^\circ$ & 14.6 & 0.33 &  175$\pm$17 & 81$\pm$15 & 82$\pm$6  \\
     &0.5   & 29$\times$27, 41$^\circ$ & 11.0 & 0.37 &  176$\pm$18 & 71$\pm$11 & 71$\pm$4 \\
     &1	    & 42$\times$34, 47$^\circ$ & 8.2  & 0.65 &  196$\pm$19 & 57$\pm$8 & 49$\pm$12  \\
     &2	    & 47$\times$40, 63$^\circ$ & 6.2  & 0.82 &  193$\pm$19 & 71$\pm$6 & 60$\pm$8\\
\hline
\hline
LB19+SB16 & -0.5& 20$\times$20, 26$^\circ$ & 26.3 & 0.27 & 184$\pm$18 & 95$\pm$26 & 97$\pm$12 \\
          & -0.3& 21$\times$21, 2$^\circ$  & 22.1 & 0.27 & 186$\pm$19 & 88$\pm$22 & 90$\pm$11\\
          & 0	& 24$\times$23, 30$^\circ$ & 15.7 & 0.30 & 188$\pm$19 & 86$\pm$16 & 89$\pm$8  \\
          &0.3  & 29$\times$26, 40$^\circ$ & 10.1 & 0.37 & 189$\pm$19 & 82$\pm$10 & 84$\pm$6 \\
          &0.5  & 36$\times$30, 44$^\circ$ & 8.8  & 0.49 & 190$\pm$19 & 80$\pm$9  & 80$\pm$8\\
          &1    & 51$\times$44, 63$^\circ$ & 4.8  & 0.96 & 189$\pm$19 & 81$\pm$5  & / \\
          &2	& 60$\times$54, 96$^\circ$ & 3.9  & 1.37 & 189$\pm$19 & 189$\pm$4 &  / \\
\hline
\hline
IB17+SB16 & -0.5& 59$\times$43, 59$^\circ$ & 24.7 & 1.01 & 176$\pm$18 & 105$\pm$25 & 111$\pm$25 \\
          & -0.3& 60$\times$44, 59$^\circ$ & 20.5 & 1.06 & 176$\pm$18 & 91$\pm$20  & 98$\pm$23  \\
          & 0	& 64$\times$48, 61$^\circ$ & 15.5 & 1.20 & 176$\pm$18 & 107$\pm$15 &  100$\pm$28\\
          &0.3  & 70$\times$54, 63$^\circ$ & 11.0 & 1.45 & 176$\pm$18 & 178$\pm$11 & 182$\pm$34 \\
          &0.5  & 75$\times$59, 64$^\circ$ & 9.1  & 1.68 & 177$\pm$18 & 264$\pm$9  &  428$\pm$38 \\
          &1    & 87$\times$69, 66$^\circ$ & 6.3  & 2.22 & 178$\pm$18 & 519$\pm$6  & 714$\pm$43\\
          &2	& 93$\times$74, 67$^\circ$ & 5.0  & 2.48 & 178$\pm$18 & 683$\pm$5  & 817$\pm$50  \\
\hline
\hline
LB19+IB17+SB16 &-0.5   & 24$\times$23, 45$^\circ$  & 16.7 & 0.29 & 173$\pm$17 & 95$\pm$17   &  95$\pm$8 \\
               &-0.3   & 26$\times$24, 41$^\circ$  & 12.9 & 0.30 & 173$\pm$17 & 89$\pm$13   &  87$\pm$6  \\
               &0      & 31$\times$26, 44$^\circ$  & 10.1 & 0.36 & 173$\pm$17 & 86$\pm$10   & 79$\pm$8 \\
               &0.3    & 40$\times$32, 47$^\circ$  & 8.4  & 0.53 & 174$\pm$17 & 83$\pm$8    & 73$\pm$9 \\
               &0.5    & 45$\times$37, 51$^\circ$  & 6.6  & 0.68 & 174$\pm$17 & 79$\pm$7    &  67$\pm$9\\
               & 1     & 63$\times$54, 78$^\circ$  & 4.4  & 1.33 & 176$\pm$18 & 170$\pm$4   &  / \\
               & 2     & 70$\times$63, 81$^\circ$  & 3.5  & 1.68 & 176$\pm$18 & 257$\pm$3   &  / \\\hline
\end{tabular}
\label{tab:imagingsummary}
\end{table*}

\begin{table*}[!h]
\centering
\caption{Extended flux in cavity, from r=1 images. The inner disk properties were derived using a Gaussian fit in an elliptical mask centered in the central pixel, sized 0.15\arcsec{}$\times$0.12\arcsec{}. Deconvolved major, minor axis FWHM and position angle are given. For the material around PDS70\,b, we defined a rectangular aperture of 0.08\arcsec{}$\times$0.15\arcsec{}, with PA=55$^\circ$.  }
\begin{tabular}{c||ccc||cc}
\hline
\multicolumn{1}{c}{{\bf Obs.}} & \multicolumn{3}{c}{{\bf Inner disk}} & \multicolumn{2}{c}{{\bf Emission around PDS70\,b}} 	\\

Dataset &  Major/minor axis, PA &   Peak $I_{\nu}$  & Integrated flux  &  Peak $I_{\nu}$  & Integrated flux\\
        & [mas$\times$mas]      &  [$\mu  \mathrm{Jy}$ $\mathrm{beam}^{-1}$ ] & [$\mu$Jy] & [$\mu  \mathrm{Jy}$ $\mathrm{beam}^{-1}$ ] & [$\mu$Jy] \\
\hline
\hline
LB19           & 129$\pm$11/93$\pm$8, 148$\pm$11$^\circ$   &  75$\pm$6  & 719$\pm$60 &  /  & /  \\
LB19+SB16      & 128$\pm$11/94$\pm$9, 152$\pm$12$^\circ$   & 126$\pm$9  & 817$\pm$69 &  46$\pm$5 & 101$\pm$10 \\
IB17+SB16      & 102$\pm$12/81$\pm$13, 171$\pm$32$^\circ$  & 367$\pm$18 & 888$\pm$59 &  56$\pm$6 & 40$\pm$4 \\
LB19+IB17+SB16 & 117$\pm$5/91$\pm$4, 166$\pm$8$^\circ$     & 174$\pm$5  & 727$\pm$27 &  27$\pm$4  & 38$\pm$3\\\hline
\end{tabular}
\label{tab:innerdisk}
\end{table*}

\subsection{Peak intensity of the continuum emission associated with the planets}
\label{sec:cpdflux}
\begin{figure}[t]
\begin{center}
\includegraphics[width=0.75\textwidth]{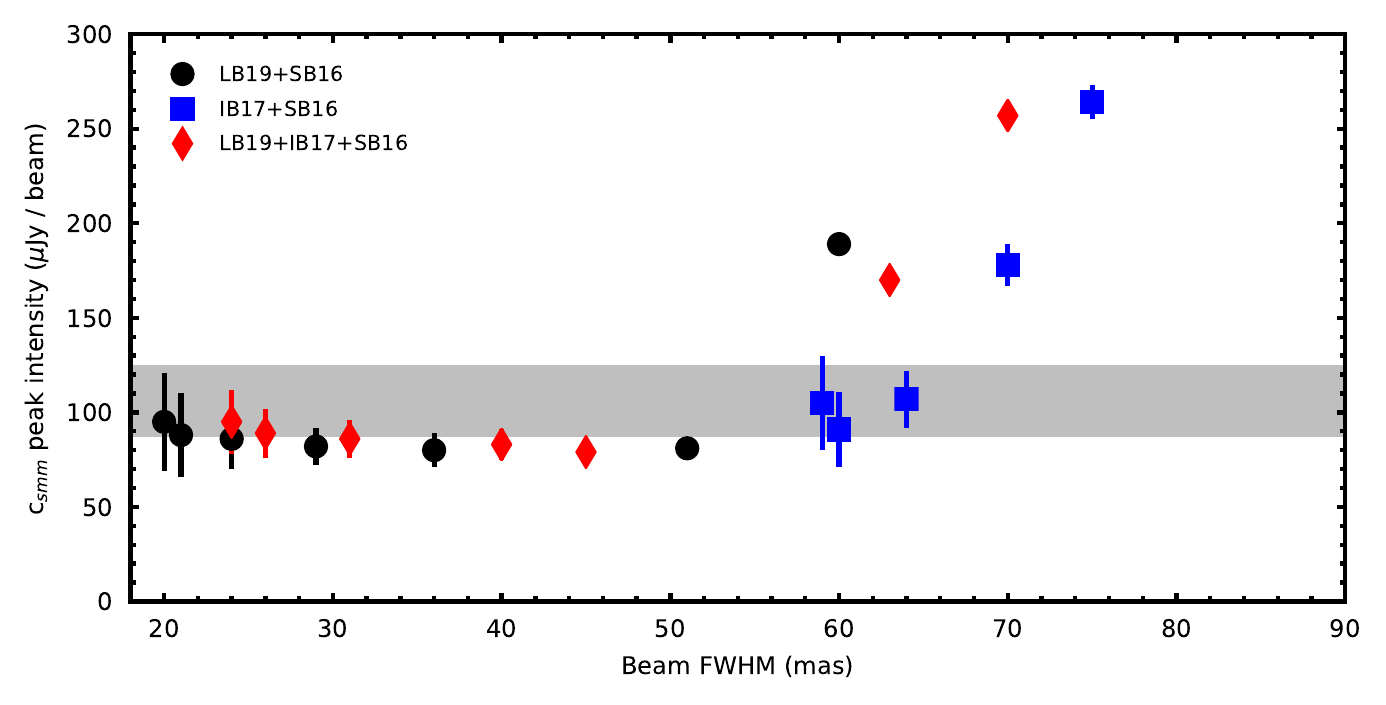} 
\end{center}
\caption{Peak intensity of c$\rm_{smm}$ as a function of angular resolution. The peak intensity of the CPD is constant around $\sim$80-105\,$\mu \mathrm{Jy}$ $\mathrm{beam}^{-1}$  when it is well separated from the outer ring. At larger resolution than $\sim$60\,mas, the peak intensity increases because the beam contains contribution from the outer disk. The grey area reports the estimate of \citet{isella2019}.} 
\label{fig:cpdflux}
\end{figure}

Figure \ref{fig:cpdflux} shows the peak intensity of c$\rm_{smm}$ as a function of angular resolution. Depending on the dataset, and the robust parameter, its peak intensity ranges between 80$\pm$6 and 107$\pm$15\,$\mu \mathrm{Jy}\mathrm{beam}^{-1}$ when it is well separated from the outer ring. At larger resolution than $\sim$60\,mas, the peak intensity increases because the beam contains contribution from the outer disk. The grey area reports the estimate of \citet{isella2019}. In contrast the peak intensity of b$\rm_{smm}$ varies between 46$\pm$5, 56$\pm$6 and 27$\pm$4\,$\mu  \mathrm{Jy}$ $\mathrm{beam}^{-1}$ for three different datasets (LB19+SB16; IB17+SB16; LB19+IB17+SB16, respectively) imaged at resolutions of  51\,mas$\times$44\,mas, 87\,mas$\times$69\,mas, 63\,mas$\times$54\,mas, respectively (see Table\,\ref{tab:innerdisk}).

\section{Outer disk visibility modelling}
\label{sec:frank}
%\twocolumngrid
To better assess the presence of substructures in the outer disk, we fit azimuthally averaged deprojected visibilities using the \texttt{frank} package that models an axisymmetric surface density profile using a flexible Gaussian process  \citep{jennings2020}. To do so, we considered the combined dataset LB19+IB17+SB16, which has the best uv coverage, assuming that the outer disk brightness distribution has not changed between these observations. The data was averaged into intervals of $30\,$s and 1 channel per spectral window to reduce data volume. 

As the disk presents an asymmetric arc-like feature in the North-West that can lead to overestimate the disk radial intensity profile when fitting with an axisymmetric model, we followed the procedure developed in \cite{andrews2021} that modifies the visibilities by removing a model for the asymmetry before fitting with \texttt{frank}. We mask the emission between position angles of  -85$^\circ$ and 40$^\circ$. The asymmetry is selected in the \texttt{CLEAN} model image, and the mean radial profile of the \texttt{CLEAN} model from the disk outside the asymmetric region is subtracted from the model image, allowing us to obtain a model for the asymmetry only. The Fourier transform of the asymmetry model was then subtracted from the original visibilities and the resulting set of visibilities are further modelled. \texttt{frank} fits deprojected visibilities, and inaccurate estimates of the geometric parameters for the deprojection, ($\Delta$RA, $\Delta$Dec, inc, PA), can lead to significant residuals. Automatic procedures performed poorly to find the best parameter set and we therefore optimized those parameters by hand, exploring different values of spatial offset and geometry in steps of 1\,mas and $0.5\,$deg. The final values adopted for the \texttt{frank} fit are (12\,mas, 15\,mas) for (dRA, dDec), and (49.5, 161.0) for (inc, pa). We tested the sensitivity of the fit to the hyperparameters $\alpha$ (varied between $1.05$ and $2.00$) and $w_{\text{smooth}}$ (varied between $10^{-4}$ to $10^{-1}$) and found no significant difference on the residuals. We therefore fixed for standard values $w_{\text{smooth}}=10^{-4}$ and $\alpha$=1.30. 

\begin{figure*}[!h]
\begin{center}
\begin{tabular}{c}
\includegraphics[width=1.0\textwidth]{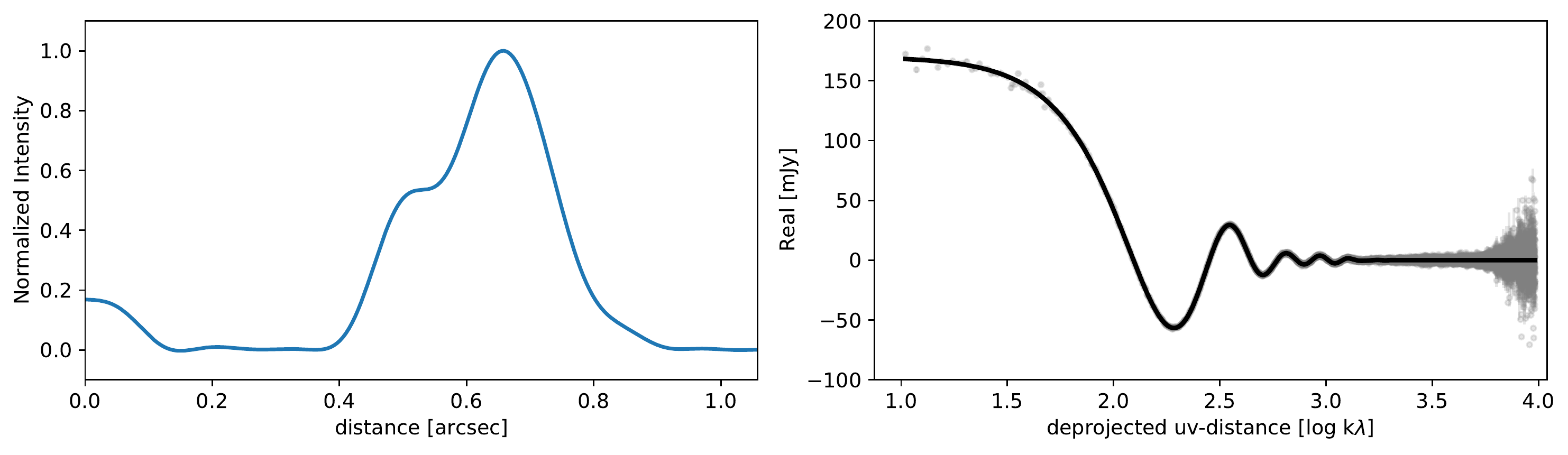}\\
\includegraphics[width=1.0\textwidth]{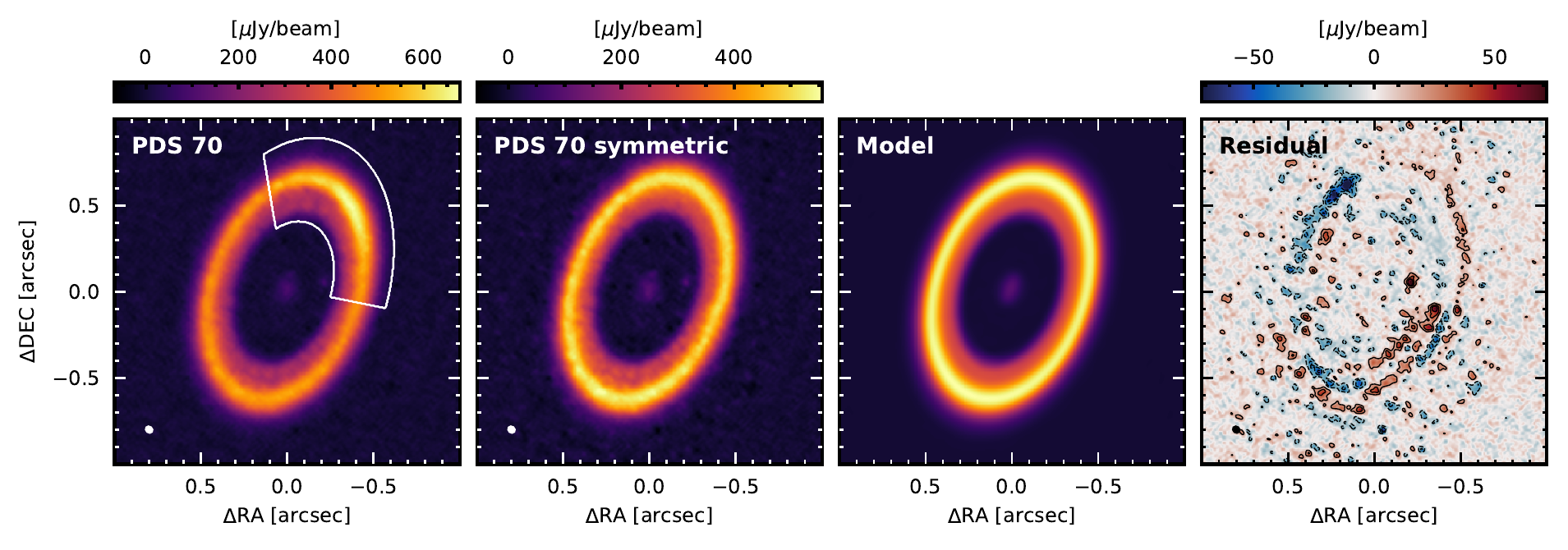}
\end{tabular}
\end{center}
\caption{Results of the 1-D modelling of the disk, with the LB19+IB17+LB16 dataset. Upper left: best radial intensity profile obtained with the \texttt{frank} package. Upper right: observed visibilities (grey) compared to the best fit model (black). Bottom panels: Left: image of the LB19+IB17+LB16 continuum emission with the mask used to build the model for the asymmetry; Middle left: the resulting symmetric continuum emission after subtracting the asymmetry; Middle right: the best \texttt{frank} model; Right: the residual map. All images are obtained with  a robust parameter of 0.5.}
\label{fig:frank_fit_profile}
\end{figure*}

The fit to the data and the corresponding profile are shown in Figure~\ref{fig:frank_fit_profile} (top panels). The best fit model indicates a radial profile with two local maxima for the emission of the outer disk, confirming the findings of \citet{Keppler2019} with lower resolution observations. At the angular resolution of our observations, the two peaks are separated by $\sim$7 beams. It is however unclear whether the outer disk is composed of two separated, broad, rings, or of a ring with an inner shoulder. No clear gap or ring is evidenced in the inner disk. We note the presence of a possible additional shoulder at 0.85\arcsec{}. 
The model and residuals are imaged exactly as the observations, with a robust parameter of 0.5, and are presented in Figure~\,\ref{fig:frank_fit_profile}, bottom panels. The residuals indicate that the axisymmetric model does not account for the full complexity of the emission, in particular the disk can be vertically thick and flared.  %The non-axisymmetric emission is clearly present in the residuals, such as the azimuthal asymmetry present in the North and c$\rm_{smm}$ and might affect the resulting fit for the geometric parameters when minimizing the imaginary part of the visibilities leading to inaccurate estimates of ($\delta$\,RA, $\delta$\,Dec, i, PA). It seems in particular that the inclination is over-estimated. 
A dedicated 2D modeling of the visibilities, that is beyond the scope of this Letter, is needed to properly assess the morphology of the disk and will be presented in a forthcoming study. %It is also possible that the geometric parameters used for the deprojection ($\delta$\,RA, $\delta$\,Dec, i,PA) are inaccurate, in particular that an inclination of 51.7$^\circ$ is slightly overestimated. 

% Following Appendix B of \citet{isella2019}, we then determine the offset ($\delta$\,RA, $\delta$\,Dec) that minimizes the imaginary part of the visibilities using a Monte Carlo Markov Chain with 32 walkers and a flat prior over both variables. We centered the visibilities on this offset and deprojected them using an inclination of 51.7$^\circ$ and a position angle of 160.4$^\circ$ \citep{Keppler2019}. 

\section{Astrometry}
\label{sec:astrometry}
Table\,\ref{tab:astrometry} provides the published astrometry of PDS\,70\,b and PDS\,70\,c, as well as the predicted locations  at the time of our observations based on the best orbital fits by \citet{wang2021}. 

\begin{table*}[!h]
\caption{Relative astrometry of PDS70 b and c.}
\label{tab:astrometry}
\begin{center}
\begin{tabular}{ccccccc}
\hline
Date & Instrument/$\lambda_0$ ($\mu$m) & $\Delta$RA (mas)  & $\Delta$Dec  (mas)  & Sep. (mas)  & PA (deg)  & Reference \\
\hline
\multicolumn{7}{c}{{\bf Astrometry of PDS70 b}} \\
2012-03-31 & NICI/L'/3.80 & 58.7$\pm$10.7 & -182.7$\pm$22.2 & 191.9$\pm$21.4 & 162.2$\pm$3.7 & \citet{Keppler2018} \\
2015-05-03 & SPHERE/H3/1.67 & 83.9$\pm$3.6 & -178.5$\pm$4.0  &197.2$\pm$4.0 & 154.9$\pm$1.1 & \citet{Keppler2018} \\
2015-05-31 & SPHERE/H2/1.59 & 89.4$\pm$6.0 & -178.3$\pm$7.1 & 199.5$\pm$6.9  & 153.4$\pm$1.8 & \citet{Keppler2018} \\
2016-05-14 & SPHERE/K1/2.11 & 90.2$\pm$7.3 & -170.8$\pm$8.6 & 193.2$\pm$8.3 & 152.2$\pm$2.3  & \citet{Keppler2018} \\
		   &				 & 	86.2$\pm$5.4  & -164.9$\pm$6.6 & 186.1$\pm$7.0 & 152.4$\pm$1.5 & \citet{haffert2019} \\
2016-06-01 & NACO/L'/3.80 & 94.5$\pm$22.0 & -164.4$\pm$27.6 & 189.6$\pm$26.3 & 150.6$\pm$7.1 & \citet{Keppler2018} \\
		   &				 & 	86.7$\pm$7.3 	& 	-159.1$\pm$9.3 			& 181.2$\pm$10.0 & 151.4$\pm$2.0 & \citet{haffert2019} \\
2018-02-24 & SPHERE/K1/2.11 & 109.6$\pm$7.9 & -157.7$\pm$7.9 & 192.1$\pm$7.9 &  147.0$\pm$2.4 & \citet{mueller2018} \\
2018-06-20 & MUSE/H$\alpha$/0.656 & 96.8$\pm$25.9  &  -147.9$\pm$25.4 			& 176.8$\pm$25.0 & 146.8$\pm$8.5 & \citet{haffert2019} \\
2019-06-08 & NIRC2/L'/3.78 &  --      & --  & 175.8$\pm$6.9 & 140.9$\pm$2.2  & Wang et al. (2020b) \\
2019-07-16 & GRAVITY/K/2.2 &  102.61$\pm$0.09   & -139.93$\pm$0.24  & -- & --  & \citet{wang2021}  \\
2020-02-10 & GRAVITY/K/2.2 &  104.70$\pm$0.09   & -135.04$\pm$0.11  & -- & --  & \citet{wang2021}\\
\hline
\hline
\multicolumn{7}{c}{{\bf Astrometry of PDS70 c}} \\
2016-05-14 & SPHERE/K1/2.11         & -207.8$\pm$6.9 & 55.7$\pm$5.7 & 215.1$\pm$7.0 & 285.0$\pm$1.5  &\citet{haffert2019} \\
2016-06-01 & NACO/L'/3.80           & -247.8$\pm$9.9 & 58.5$\pm$8.9 & 254.1$\pm$10.0 & 283.3$\pm$2.0 &\citet{haffert2019} \\
2018-02-24 & SPHERE/K12/2.2        & -205.0$\pm$13.0 & 41.0$\pm$6.0 & 209.0$\pm$13.0 & 281.2$\pm$0.5 &\citet{mesa2019} \\
2018-06-20 & MUSE/H$\alpha$/0.656   & -233.7$\pm$25.0 & 28.8$\pm$26.7 & 235.5$\pm$25.0 & 277.0$\pm$6.5 &\citet{haffert2019} \\
2019-03-06 & SPHERE/K12/2.2        & -222.0$\pm$8.0  & 39.0$\pm$4.0 & 225.0$\pm$8.0  & 279.9$\pm$0.5 &\citet{mesa2019} \\
2019-06-08 & NIRC2/L'/3.78          & --         & --  &           223.4$\pm$8.0 & 280.4$\pm$2.0  &\citet{wang2021} \\
2019-07-19 & GRAVITY/K/2.2 &  -214.95$\pm$0.13   & 32.22$\pm$0.13  & -- & --  & \citet{wang2021} \\
2020-02-10 & GRAVITY/K/2.2 &  -214.30$\pm$0.07   & 27.19$\pm$0.16  & -- & --  & \citet{wang2021} \\
\hline
\hline
% \multicolumn{7}{c}{{\bf Astrometry of PDS~70~c$_{smm}$}} \\
% 2017-12-03 & ALMA/855 cont    & -213.1$\pm$3.5 & 47.0$\pm$4.9 &      --          &        --       &  \citet{isella2019} \\  
% 2019-07-27 & ALMA/855 cont    & \todo{XX}  & \todo{XX}   &      --          &        --       &  This work \\  
% \hline
% \hline
\multicolumn{7}{c}{{\bf Predicted astrometry of PDS~70~b}} \\
2017-12-04 &  & 96.86$\pm$1.03   & -153.66$\pm$0.63  & 181.76$\pm$0.78  & 147.80$\pm$0.29 &  \citet{wang2021} \\
2019-07-28 &   & 103.69$\pm$0.99  & -139.97$\pm$0.22  &174.12$\pm$0.69   & 143.48$\pm$0.24  &  \citet{wang2021} \\
\hline
\multicolumn{7}{c}{{\bf Predicted astrometry of PDS~70~c}} \\
2017-12-04  & &-216.18$\pm$0.58   & 46.03$\pm$0.68 & 221.04$\pm$0.60  & 282.01$\pm$0.17 &  \citet{wang2021} \\
2019-07-28 & & -214.81$\pm$0.32  & 31.87$\pm$0.41 & 217.16$\pm$0.36  & 278.43$\pm$0.10   & \citet{wang2021} \\

\end{tabular}
\end{center}
\end{table*}

\section{CPD mass ranges}
\label{sec:cpdmass}
\begin{figure}[!h] 
\begin{center}
\includegraphics[width=0.5\textwidth]{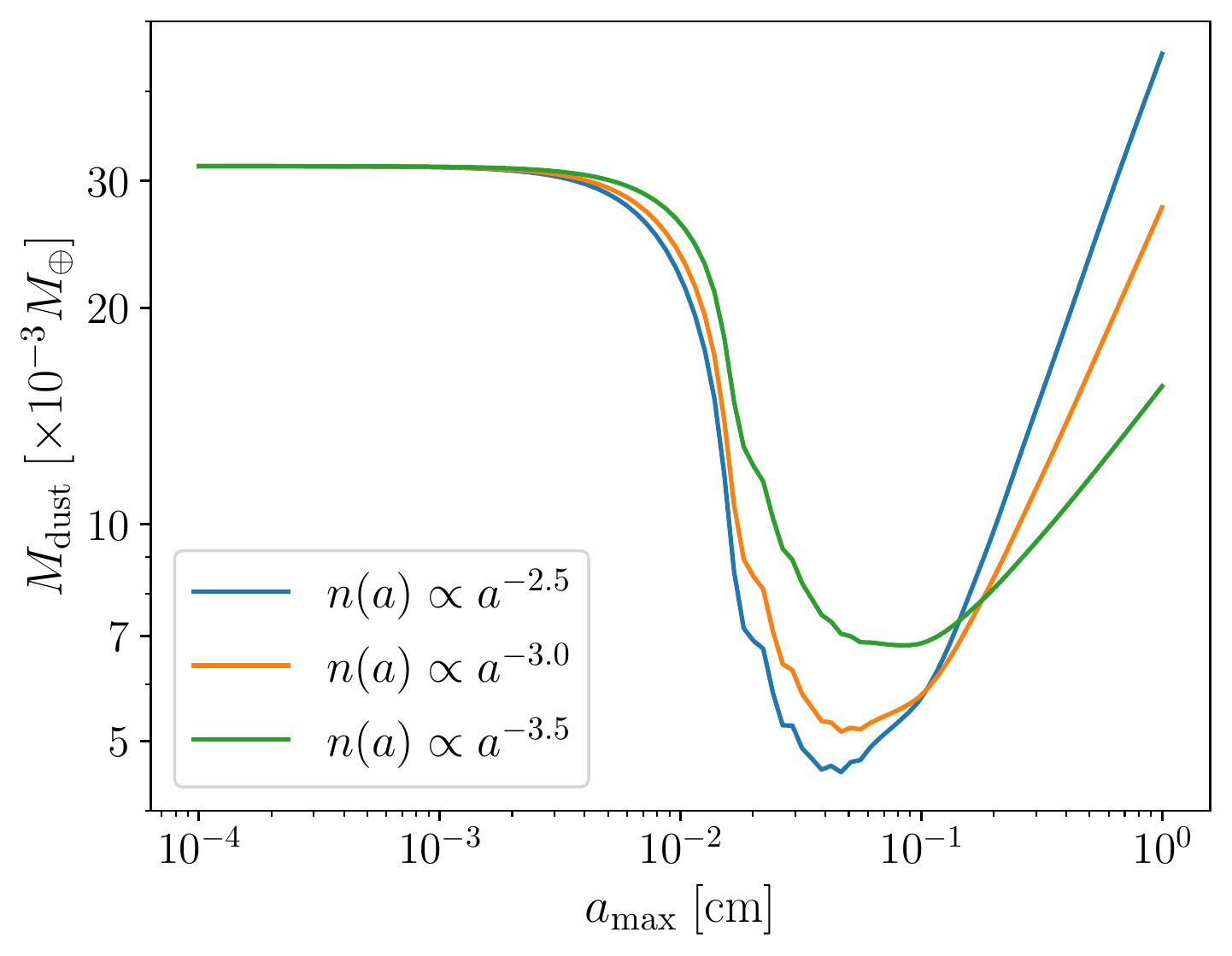}
\end{center}
\caption{Dust mass in the CPD around PDS\,70\,c for various dust grain size distribution, as a function of the maximum grain size. All the grain size distributions varies from $a_{\rm{min}}$ = 0.05\,$\mu$m to $a_{\rm{max}}$.}
\label{fig:Mcpd_amax}
\end{figure}

We consider 3 models for the dust grain population in the CPD around PDS\,70\,c, that follow different size distribution $n(a)da \propto a^{-3.5}da$,$\propto a^{-3.0}da$, and $\propto a^{-2.5}da$ and show the predicted dust mass as a function of the maximum grain size $a_{\rm{max}}$ in Figure~\ref{fig:Mcpd_amax}. We consider a minimum grain size of 0.05\,$\mu$m, a temperature of 26~K and use the DSHARP opacities \citep{Birnstiel2018}.

\acknowledgments
We are very grateful to Marco Tazzari and Antonella Natta for their support during the Covid-19 pandemic. We thank the North American ALMA ARC for their help. We acknowledge Ryan Loomis, Jason Wang and Rens Waters for insightful discussions and the reviewer for helpful suggestions that improved the manuscript. This Letter makes use of the following ALMA data: ADS/JAO.ALMA\#2018.A.00030.S.,  ADS/JAO.ALMA\#2017.A.00006.S, ADS/JAO.ALMA\#2015.1.00888.S. ALMA is a partnership of ESO (representing its member states), NSF (USA), and NINS (Japan), together with NRC (Canada),  NSC and ASIAA (Taiwan), and KASI (Republic of Korea), in cooperation with the Republic of Chile. The Joint ALMA Observatory is operated by ESO, AUI/NRAO, and NAOJ. This project has received funding from the European Research Council (ERC) under the European Union’s Horizon 2020 research and innovation programme (grant agreement No. 101002188 and No. 832428). JB acknowledges support by NASA through the NASA Hubble Fellowship grant \#HST-HF2-51427.001-A awarded  by  the  Space  Telescope  Science  Institute,  which  is  operated  by  the  Association  of  Universities  for  Research  in  Astronomy, Incorporated, under NASA contract NAS5-26555. SF acknowledges an ESO Fellowship. S.A. acknowledges support from the National Aeronautics and Space Administration under grant No. 17-XRP17 2-0012 issued through the Exoplanets Research Program. A.I. acknowledges support from the National Science Foundation under grant No. AST-1715719
and from NASA under grant No.  80NSSC18K0828. J.M.C. acknowledges support from the National Aeronautics and Space Administration under grant No. 15XRP15\_20140 issued through the Exoplanets Research Program. N.T.K. and P.P. acknowledge support provided by the Alexander von Humboldt Foundation in the framework of the Sofja Kovalevskaja Award endowed by the Federal Ministry of Education and Research.

\vspace{2mm}
\facilities{ALMA}
\software{\texttt{GALARIO} \citep{galario}, \texttt{CASA} \citep{McMullin2007}, \texttt{Matplotlib} \citep{matplotlib}, \texttt{numpy} \citep{numpy}, \texttt{emcee} \citep{emcee}, \texttt{frank} \citep{jennings2020}}

\bibliography{pds70}{}
\bibliographystyle{aasjournal}

%TC:endignore
\end{document}